\documentclass[
 reprint,
superscriptaddress,
 amsmath,amssymb,
 aps,
prb,
]{revtex4-2}

\usepackage{graphicx}
\usepackage{dcolumn}
\usepackage{bm}
\usepackage[version=3]{mhchem}
\usepackage{blindtext}
\usepackage{mathtools}
\usepackage{siunitx}
\usepackage{makecell}
\usepackage{xfrac}

\usepackage[normalem]{ulem}

\usepackage{hyperref}
\hypersetup{colorlinks=true,citecolor=blue,linkcolor=blue,filecolor=blue,urlcolor=blue,pdfstartview=FitH,pdfpagemode=UseNone}
\usepackage{xcolor}

\newcommand{\cdash}{\mathrel{\raisebox{0.6ex}{\rule{0.4em}{0.3pt}}}}

\newcommand{\dIdV}{$\mathsf{d}I/\mathsf{d}V\!\!\cdash\!$ }

\newcommand{\CO}{\ce{CO}$\cdash$}

\begin{document}

\preprint{APS/123-QED}

\title{Probing the Electronic Structure at the Boundary of Topological Insulators in the \ce{Bi2Se3} Family by Combined STM and AFM}

\author{Christoph S. Setescak}
\email{Christoph.Setescak@physik.uni-regensburg.de}
\affiliation{%
 Institute of Experimental and Applied Physics, University of Regensburg, Universitätstraße 31, 93053 Regensburg, Germany
}
\author{Irene Aguilera}
\affiliation{%
Institute for Theoretical Physics, University of Amsterdam and European Theoretical Spectroscopy Facility (ETSF), Science Park 904, 1098 XH Amsterdam, The Netherlands
}
\author{Adrian Weindl}
\affiliation{%
 Institute of Experimental and Applied Physics, University of Regensburg, Universitätstraße 31, 93053 Regensburg, Germany
}
\author{Matthias Kronseder}
\affiliation{%
 Institute of Experimental and Applied Physics, University of Regensburg, Universitätstraße 31, 93053 Regensburg, Germany
}
\author{Andrea Donarini}
\affiliation{%
 Institute of Theoretical Physics, University of Regensburg, Universitätstraße 31, 93053 Regensburg, Germany
}
\author{Franz J. Giessibl}
\affiliation{%
 Institute of Experimental and Applied Physics, University of Regensburg, Universitätstraße 31, 93053 Regensburg, Germany
}


\date{\today}

\begin{abstract}
We develop a numerical scheme for the calculation of tunneling current $I$ and differential conductance $\mathsf{d}I/\mathsf{d}V$ of metal and \mbox{\CO terminated} STM tips on the topological insulators \ce{Bi2Se3}, \ce{Bi2Te2Se} and \ce{Bi2Te3} and find excellent agreement with experiment. The calculation is an application of Chen's derivative rule, whereby the Bloch functions are obtained from Wannier interpolated tight-binding Hamiltonians and  maximally localized Wannier functions from first-principle DFT+$GW$ calculations. We observe signatures of the topological boundary modes, their hybridization with bulk bands, Van Hove singularities of the bulk bands and characterize the orbital character of these electronic modes using the high spatial resolution of STM and AFM. Bare DFT calculations are insufficient to explain the experimental data, which are instead accurately reproduced by many-body corrected $GW$ calculations.
\end{abstract}

\maketitle


\section{Introduction}
\label{sec:i}

Solid-state physics seeks to understand the large-scale properties of materials, in principle deducing it from  atomic- scale behavior. 
In insulators, the atomic arrangement and electron-electron interaction results in bulk valence and conduction bands that are separated by an energy gap.
These bands are classified using topological invariants, which define distinct topological phases of matter \cite{Thouless1982, Kane2005, Kane2005_2, Fu2007}. 
In three-dimensional topological insulators, non-trivial invariants give rise to gapless boundary modes, linking atomic-scale behavior to emergent large-scale phenomena. These boundary modes are characterized by spin-polarized, nearly linear dispersion relations, thus effectively describing massless particles.

Prototypical examples of three-dimensional topological insulators include materials from the \ce{Bi2Se3} family, such as \ce{Bi2Se3}, \ce{Bi2Te2Se}, and \ce{Bi2Te3}. 
The dispersion relation 
of topological insulators can be very well studied using angle resolved photoemission spectroscopy (ARPES)~\cite{Lv2019}. 
However, standard ARPES is only able to probe occupied states and more advanced time-resolved ARPES (trARPES) is necessary to probe unoccupied states~\cite{Lv2019}.
Furthermore, ARPES is limited to the momentum-space and to length scales much larger than the atomic ones.
Transport and ferromagnetic resonance experiments have demonstrated that atomic scale defects have a large influence on sample properties \cite{Knispel2017, Mayer2021, Pietanesi2024}.

To overcome these limitations and explore the atomic-scale phenomena in real space, scanning probe microscopy techniques, such as scanning tunneling microscopy (STM), atomic force microscopy (AFM), and scanning tunneling spectroscopy (STS), are helpful.

Examples of previous scanning probe studies on materials in the \ce{Bi2Se3} family are plentiful \cite{Urazhdin2004, Hor2009, Romanowich2013}. 
The influence of native point defects on the local density of states ($\sf{LDOS}$) in \ce{Bi2Te3} has been documented in Ref.~\cite{Jurczyszyn2020} and impurity resonances in \ce{Bi2Se3} have been studied in Ref.~\cite{Yeh2012}. Furthermore
it is possible to identify the doping character of native point defects \cite{Jurczyszyn2020, Netsou2020}, which can also be probed directly via AFM \cite{Liebig2022}.
Scattering at step edges on \ce{Bi2Te3} has been used in Ref.~\cite{Alpichshev2010} to quantify the dispersion and verify the spin polarization of the topological boundary mode. 
In Refs.~\cite{Dmitriev2014, Fedotov2017} changes in the $\sf{LDOS}$ close to few-atomic steps on \ce{Bi2Se3} have been studied, which indicate increased conductance in the energy range of the topological boundary mode close to the step edge. 
The necessity of quantitatively accurate calculations of the \dIdV signal becomes apparent from this work, as such changes can be caused and explained by changes in the local work function and chemical potential close to step edges. 
This is in contrast with other possible explanations, such as additional one-dimensional electronic states.
Further, recent theoretical studies propose atomic-scale experiments that would use ring states at defects as a probe for bulk topological properties \cite{Queiroz2024}.

In this article we wish to establish how the band structure of these materials can be studied using experimental techniques that access the atomic-scale.
New experimental results presented here include measurements of the \dIdV signal over a bias range where characterizing bulk energy levels is possible. 
Furthermore, we probe the \dIdV signal throughout the valence and conduction bands with atomic resolution, giving us a grip on characterizing the orbital character of the Bloch bands.

We bridge the gap between the atomic scale and the macroscopic properties, such as the band structure, which relies on translational symmetry of a large crystal.
State-of-the-art $GW$ calculations 
are used to construct tight-binding Hamiltonians for thick slab geometries and are the basis for calculating the differential conductance $\mathsf{d}I/\mathsf{d}V$ through the atomic-scale tip-sample junction.
Due to the large sample geometries in our experiment, \emph{i.e.}, semi-infinite single crystals or thick slabs, it is not feasible to study these directly with $GW$ or DFT, necessitating the use of the TB Hamiltonians.
Employing $GW$ calculations is an improvement over previous studies, as it has been shown that bare DFT calculations are not sufficient to accurately describe this material class~\cite{Aguilera2019}.
In our calculation we include the energy dependence of the \dIdV signal as well as the dependence on the position of the tip apex $\bold{r}_{0}$.

The paper is organized as follows. 
In Section~\ref{sec:intro} we give an overview of our findings and review relevant literature.
Our methods are described in detail in Section~\ref{sec:methods}.
The results of scanning tunneling spectroscopy on the three compounds are in Section~\ref{sec:STS}. 
Section~\ref{sec:mbc} serves to show how these results can be applied to probe the influence of many-body corrections on the band structure.
Section~\ref{sec:orbital} reports how the orbital character of the band structure can be probed using our experimental setup.
Our conclusions are summarized in Section~\ref{sec:out}

\section{Overview and Comparison to Previous Results}
\label{sec:intro}
\begin{figure*}
        \centering
        \includegraphics{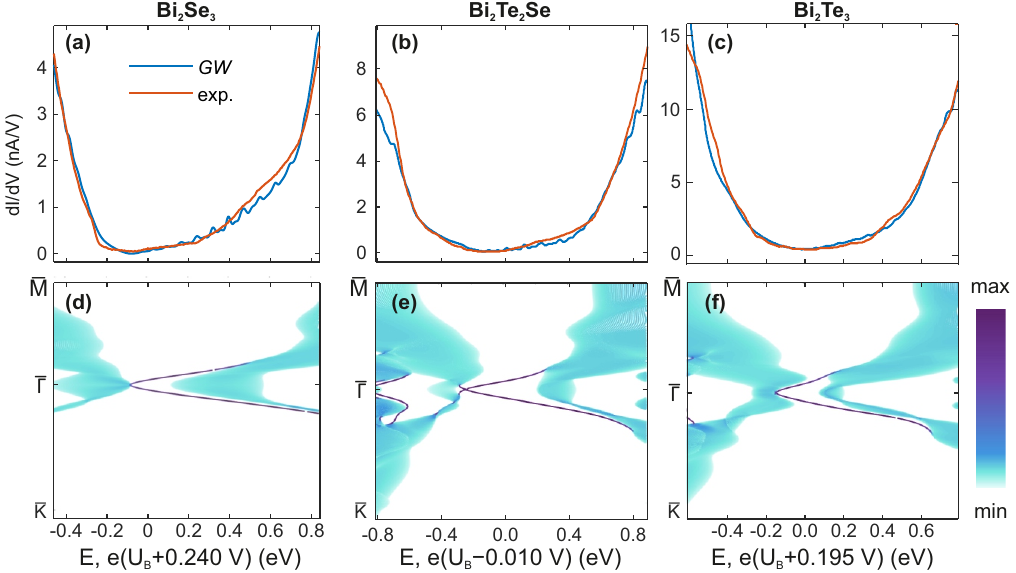}
        \caption{\label{fig:intro} Comparison of simulated and experimental \dIdV spectrum of a \CO functionalized tips on the (a) surface of \ce{Bi2Se3}, (b) surface of \ce{Bi2Te2Se} and (c) surface of \ce{Bi2Te3}. The half-space band structures are plotted in (d) - (f). The energy axes for each compound are identical and are related to the bias voltage by a shift, due to defects doping the crystal. The colorscale in (d)-(f) indicates the fraction of the eigenstate in the first quintuple layer.} 
\end{figure*}
Figure \ref{fig:intro} shows the main result of this article. Panels \ref{fig:intro} (a) to (c) display in orange \dIdV spectra measured with \CO terminated tips on the surfaces of \ce{Bi2Se3}, \ce{Bi2Te2Se}, and \ce{Bi2Te3}, alongside the calculated  \dIdV spectra in blue. 
The corresponding half-space band structures are presented in Figure \ref{fig:intro} (d) to (f).
Here, the colorbar indicates the projection of the states onto the Wannier functions associated to the atoms of the topmost quintuple layer (QL), with bulk bands appearing in light blue and surface states shown as dark purple lines. 
In the following we want to point out thirteen observations.

(\emph{i}) Our numerical scheme for calculating the \dIdV spectra can be summarized in five steps.
Starting point are Wannier interpolated tight-binding (TB) Hamiltonians for materials in the \ce{Bi2Se3} family derived from many-body renormalized $GW$ calculations \cite{Aguilera2019}.
From these we construct half-space Hamiltonians using Dirichlet boundary conditions.
We represent the Bloch functions in real-space using maximally localized Wannier functions \cite{Marzari2012}.
Propagation of the Bloch functions into the tip-sample junction is modeled under the assumption of non-interacting electrons traveling in a laterally averaged potential. 
Eventually, the electron tunneling into and out of the metal and \CO terminated tips is calculated using Chen's derivative rule \cite{Bardeen1961, Chen1990}. 

(\emph{ii}) Metal tips are typically preferred for \dIdV measurements. They have a simpler apex electronic structure dominated by an $s$-wave orbital, whereas a linear combination of $s$- and $p$-wave orbitals is necessary to properly describe \CO terminated tips. 
With the calculation method presented in Section \ref{sec:methods} and in particular Subsection~\ref{ssec:cotc} it is possible to accurately model the tunneling matrix element into a mix of tip orbitals. This increased complexity of the \CO terminated tip in STS experiments is outweighed, though, by their very high chemical stability and high spatial resolution, as demonstrated for example in Section \ref{ssec:spatm}. For a better comparison to existing literature, we also show the spectrum of a metal tip on the surface of \ce{Bi2Se3} in Section \ref{Bi2Se3_STS}.

(\emph{iii}) Throughout the article we give the energy $E$ relative to the calculated Fermi level~\cite{Aguilera2019}.
Due to defects in crystal samples, the experimental Fermi level is shifted by $\Delta E_F$.
The experimental counterpart of the energy $E$ is the bias voltage $U_B$. To facilitate a comparison between experiment and theory via the plots, we shift the bias voltage $U_B$ by $\Delta E_F$. The magnitude of $\Delta E_F$ is a measure for the level of unintentional doping in the crystals.

(\emph{iv}) The general shape of the \dIdV spectrum is similar across the three compounds, as can be seen in Figure~\ref{fig:intro} (a)-(c). 
In the region close to the bulk band gap the differential conductance is greatly reduced.
Here we see linear relations between the \dIdV signal and bias voltage $U_B$.
Towards the bulk bands the \dIdV signal increases rapidly.
We find excellent agreement between experiment and theory and give an overview of the findings in the remainder of this section.
A more detailed discussion follows in Sections~\ref{sec:STS}, \ref{sec:mbc} and \ref{sec:orbital}.


(\emph{v}) The emergence of the gap-closing boundary mode upon restricting the system to half-space has been studied since the first band structure calculations of three dimensional topological insulators, published in Ref.~\cite{Zhang2009}. 
The signature of this boundary mode in the \dIdV spectrum is a linear increase of the differential conductance inside the bulk band gap.
Focusing on the compound \ce{Bi2Se3}, this linear increase can be seen in the experimental data and calculation in Figure~\ref{fig:intro} (a) for $\SI{-0.2}{eV} < E < \SI{0.23}{eV}$.
For a single band with index $n$ and dispersion relation $E_n(\bold{k})$ the partial density of states is given as
\begin{equation}
    \label{eq:ldos}
    \begin{split}
        \sf{DOS}_n(\bold{r}, E) 
        &= \frac{1}{(2\pi)^2}\int_{\sf{BZ}}  \delta{(E - E_n(\bold{k}))} \, \rm{dk}_x\rm{dk}_y\\
        &=\frac{1}{(2\pi)^2} \int_{\sf{CEC}_n(E)}  \|\mathrm{grad}(E_n)(\bold{k})\|^{-1} \, \rm{dS}.
    \end{split}
\end{equation}
Here $\mathsf{CEC}_n(E)$ is the constant energy contour of the boundary mode $n$ at energy $E$, \emph{i.e.} the curve consisting of all $\bf{k} \in \mathsf{BZ}$ with $E_n(\mathbf{k}) = E$. For a linear and isotropic dispersion relation, we have $\| \mathrm{grad}(E_n)(\bold{k})\| = \hbar \mathit{v}_F$, where $v_F$ is a constant referred to as the Fermi velocity \cite{Ashcroft1976}. The constant energy contour $\sf{CEC}_n(E)$ of such a topological boundary mode is circular and the line integral in Equation \eqref{eq:ldos} can be evaluated to be 
\begin{equation}
    \label{eq:sldos}
    \mathsf{DOS}_n(\bold{r}, E) = \frac{\vert E-E_D\vert}{2 \pi \hbar^2 v_F^2},
\end{equation}
where $E_D$ is the energy of the Dirac point.
This result follows from calculating the circumference of the $\mathsf{CEC}_n(E)$, which is equal to $\frac{2\pi}{\hbar v_F}\vert E -E_D\vert$.
The differential conductance $\mathsf{d}I/\mathsf{d}V$ is closely related to the $\sf{DOS}$, and thus from the Relation \eqref{eq:sldos} it becomes clear that we expect a linear increase of the \dIdV signal away from the Dirac point within the band gap.
This linear section 
is reproduced by the more involved quantitative calculation introduced in Section~\ref{sec:methods}.
We note that this key signature of topological boundary modes can also be explained by the simplified tight-binding model described in Refs.~\cite{Zhang2009, Leung2012}, which only considers four electronic degrees of freedom. 

(\emph{vi}) In the experiment we also observe hybridizations of the surface state with the bulk bands. 
Among others, this hybridization with the bulk bands is important for describing transport experiments \cite{Lewenkopf2018}. 
To properly describe the hybridization of the boundary mode with the bulk modes, a framework beyond an effective, low-energy tight-binding model is required.
This hybridization is particularly interesting for the boundary mode of \ce{Bi2Se3} when it intersects the valence band edge.
From Figure~\ref{fig:intro} (a) and (d), it can be deduced that the minimum of the \dIdV spectrum on \ce{Bi2Se3} corresponds to the Dirac point, which agrees with previous studies \cite{Dmitriev2014}.
The linear increase in differential conductance away from the Dirac point \( (E_D = \SI{-0.14}{eV}) \), as shown in Figure \ref{fig:intro} (a), was attributed to contributions from the upper and lower halves of the Dirac cone ~\cite{Dmitriev2014}.
This would be a fitting interpretation, if \ce{Bi2Se3} were a, possibly indirect, semiconductor with the Dirac point isolated from the bulk bands.
However, the $GW$ calculations, ARPES, and magneto-optical measurements~\cite{Nechaev2013,Orlita2015} show that \ce{Bi2Se3} is a direct band gap semiconductor and the Dirac point coincides with the valence band maximum (VBM) at the $\sf \Gamma$ point, as displayed in Figure \ref{fig:intro} (d). 
Nonetheless, our calculation, the blue curve in Figure \ref{fig:intro} (a), shows that this linear section in the \dIdV spectrum can be explained also by tunneling into the bulk valence bands.
Thus the attribution of the linear increase in the differential conductance for $E< E_D$ to a lower half of the Dirac cone has to be modified.
This is of great importance if one is interested in studying a gap opening in the topological boundary mode \cite{SanchezBarriga2018}.
The situation is analogous for metal tips, a detailed treatment of which is given in Section \ref{Bi2Se3_STS}.

(\emph{vii}) In our experiment, we also find signatures that allow for investigation of the bulk bands. 
We observe that the bulk modes near the conduction band edge contribute minimally to the \dIdV signal, as the tunneling matrix element for states close to the conduction band edge is significantly suppressed. 
This suppression is clearly shown for the three compounds in Section~\ref{sec:STS}. In the \dIdV spectrum measured on the surface of \ce{Bi2Se3}, shown in Figure~\ref{fig:intro} (a), sharp increases are observed at $E = \SI{-0.23}{eV}$ and $E = \SI{0.72}{eV}$. These energies correspond to a flattening of the dispersion relation in the $\overline{\mathsf{\Gamma}}$--$\overline{\mathsf{M}}$ direction, visible in Figure~\ref{fig:intro} (d). Similar observations can be made for \ce{Bi2Te2Se} and \ce{Bi2Te3}.
A quantitatively correct tight-binding model, that reproduces first-principle calculations over the entire Brillouin zone and an extended energy range is necessary. 
Such models which reproduce DFT-band structures have been already calculated in Ref.~\cite{Zhang2009} and Slater-Koster parameters were published in Ref.~\cite{Kobayashi2011} for \ce{Bi2Se3} and \ce{Bi2Te3}.
However, we find that standard DFT calculations are not sufficient to correctly describe the experimental observations. 
Rather, many-body corrections are necessary, as
the flattening of the dispersion relation is closely related to the size of the trivial band gaps at the Brillouin zone boundary and the $GW$ correction has a strong influence on the size of these band gaps~\cite{Aguilera2019}. We find that the experimental observations are explained accurately by the $GW$ corrected TB Hamiltonians.
A comparison between the $GW$ corrected TB Hamiltonians and the TB Hamiltonians without many-body correction for \ce{Bi2Se3} is given in Section~\ref{sec:mbc}.
Thus, we demonstrate that not only are the drastic changes in the band structure near $\sf \Gamma$, induced by the $GW$ correction, consistent with experimental observations, but the values of the trivial band gaps far from the $\sf \Gamma$ point also align more closely with the $GW$ calculations.

(\emph{viii}) Arguably the compound \ce{Bi2Te2Se} comes closest to the idealized notion of three-dimensional topological insulators, put forth by effective TB models such as in Ref.~\cite{Zhang2009}.
For one it is an indirect semiconductor and the Dirac point lies below the VBM, evident from Figure~\ref{fig:intro}~(e). Furthermore it displays a large bulk band gap, and thus the topological boundary mode is separate from the bulk bands over a large energy range.
The experimental and calculated \dIdV spectrum is plotted in Figure~\ref{fig:intro}~(b). 
The bulk band gap and thus also the energy range on which an isolated Dirac cone can be observed is on the order of \SI{0.37}{eV}, much larger for \ce{Bi2Te2Se} than for \ce{Bi2Se3} and \ce{Bi2Te3}, as can be seen in the half-space band structure which is plotted in Figure~\ref{fig:intro}~(e). In Section \ref{Bi2Te2Se_STS} we discuss the \dIdV spectrum of a \CO terminated tip on the surface of \ce{Bi2Te2Se} with focus on the energy range around the bulk band gap. 

(\emph{ix}) Furthermore, a desirable property of the ternary compound is its much smaller level of bulk doping, compared to the binary compounds studied in this article.
This is in part due to the presence of both hole- and electron-producing defects \cite{Fuccillo2013, Mayer2021, Pietanesi2024}. 
In our study, we detected slight $n$-doping or slight $p$-doping after cleaving the \ce{Bi2Te2Se} sample in the UHV chamber.
Figure~\ref{fig:intro} (b) shows that the chemical potential is shifted by $\Delta E_F = \SI{-10}{meV}$ relative to the calculated Fermi level, while Figure~\ref{fig:dIdV_Bi2Te2Se} in Section~\ref{Bi2Te2Se_STS} presents a spectrum from a different location on the sample, revealing a shift of $\Delta E_F = \SI{10}{meV}$.
Such spatial variations are in agreement with previous studies of the surface potential fluctuations on this material class~\cite{Knispel2017, Pietanesi2024}.
The position of the Fermi level is of vital importance in transport measurements.
Our experiments are not significantly affected by the slight $n$-doping, or even $p$-doping, as we can probe unoccupied as well as occupied states. 
This is in contrast to ARPES experiments, where additional dopants have to be added to the surface in order to tune the energetic position of the Dirac point \cite{Neupane2012} or, alternatively, trARPES has to be used. 
To first approximation, a shift in the chemical potential leads to a shift of the entire band structure~\cite{Hor2009}.
But, strictly speaking, due to the screening of the electron-electron interaction, the position of the chemical potential does influence the band profile~\cite{Bruus2004}. 
As the first-principles calculations for \ce{Bi2Se3} and \ce{Bi2Te3} do not consider shifts in the chemical potential due to dopants, there is a small error. 
However, for \ce{Bi2Te2Se} the shift in the chemical potential is minimal, and overall we see a similar level of agreement between experiment and theory. 
Thus we can conclude that this effect is minimal, for the systems studied in this work.

(\emph{x}) With AFM we observe a large number of substitutional defects in the first atomic layer of \ce{Bi2Te2Se}. Presumably these are due to the chemical and physical similarities of the \ce{Te} and \ce{Se}.
We position the tip as far away from these substitutional defects as possible for all the measurements presented on \ce{Bi2Te2Se} in this article.


(\emph{xi}) Due to the very small bulk band gap of \ce{Bi2Te3} the linear section in the \dIdV spectrum shown in Figure \ref{fig:intro} (c) is hard to identify. 
The half-space band structure computed with the $GW$ Hamiltonian and plotted in Figure \ref{fig:intro} (f) indicates that an isolated Dirac cone exists only for an energy range of \SI{0.07}{eV}, the lower part of the Dirac cone being covered by the bulk valence bands.
As a result of the smaller bulk band gap, the onset of the bulk bands is energetically much closer together than for the other two compounds. 
The \ce{Bi2Te3} sample was prepared by MBE and consists of $7$ quintuple layers, which in previous work has been shown to be sufficient for compounds in this material class to behave bulk like \cite{Zhang2010}.
On the MBE sample we observe an elevated defect density compared samples cleaved from single crystals of \ce{Bi2Se3} \cite{Liebig2022}. 
The sample preparation processes are described in Section \ref{ssec:em}. 

(\emph{xii}) In previous works \cite{Urazhdin2004, Romanowich2013, Netsou2020} a similar comparison between experimental and calculated \dIdV spectra on these compounds was attempted by the means of pure first-principles calculations. 
The drawback of a full first-principles calculation is that the integration over the Brillouin zone can only be carried out on a very coarse grid of crystal momenta, due to the high computational cost. 
In particular, a $24 \times 24$ Monkhorst-Pack grid was used in Ref.~\cite{Netsou2020}. Here, due to the computational low burden of the tight-binding model we can perform the integration on a grid of $80 \times 80$ crystal momenta in the surface Brillouin zone on a personal computer with $\SI{64}{GB}$ of RAM. The result is, that numerical artefacts are virtually eliminated in the calculation of the \dIdV spectra, while we use physically realistic smoothing parameters. A drawback of our method is, that we do not consider surface effects self-consistently. However, we believe that the comprehensive agreement between experiment and theory justifies \emph{a posteriori} our approach.

(\emph{xiii}) For \ce{Bi2Te3} we present a novel result in Section~\ref{Bi2Te3_STS}, which is \dIdV spectroscopy over a large bias range of $U_B \in [\SI{-2}{V},\,\SI{2}{V}]$. 
\emph{A priori} the authors did not expect that it would be possible to reproduce this experimental result with the precision demonstrated in Figure~\ref{Bi2Se3_dIdV} of Section~\ref{Bi2Te3_STS}.
During the analysis of this data, we explain how Van-Hove singularities due to a flat dispersion of electronic modes at the Brillouin zone boundary lead to the emergence of characteristic peaks in the \dIdV spectra. 
In the past, one hurdle regarding the interpretation of \dIdV spectra has been their sometimes feature-less U-shaped appearance.
Our novel result introduces means to circumvent this problem for a more rigorous analysis.
In Section~\ref{sec:mbc} we use comparable \dIdV spectra on \ce{Bi2Se3} to explain that the experimental observation aligns with the $GW$ calculation, whereas pure DFT falls short of reproducing experimental data.
Another application is given in Section \ref{ssec:specm} at the example of \ce{Bi2Te3}. By measuring \dIdV spectra within $\SI{-2}{V} < U_B < \SI{2}{V}$ above both \ce{Te} atoms and hollow sites on the \ce{Bi2Te3} surface, the orbital character of the Bloch bands can be investigated.





\section{Methods}
\label{sec:methods}
\subsection{First-Principles Methods}
\label{ssec:fpm}
The first-principles calculations were performed with density functional theory (DFT) and the $GW$ method with the schemes and convergence parameters discussed in Ref.~\cite{Aguilera2019}. The results are based on the all-electron full-potential augmented plane-wave (FLAPW) formalism, as implemented in the DFT code FLEUR~\cite{fleur} and the $GW$ code SPEX~\cite{Friedrich2010}. For the $GW$ calculations, the spin-orbit coupling was included already in the mean-field starting point, following the approach of Ref.~\cite{Aguilera2013}. 

The $GW$ approximation includes renormalization effects due to electron-electron interaction contributions that are neglected in DFT. $GW$ has consistently shown improved results compared to DFT in the theoretical description of this family of materials. This includes a variety of properties such as critical points of topological phase transitions~\cite{SanchezBarriga2018}; band gaps, effective masses, and spin-orbit splittings~\cite{Aguilera2013_2, Aguilera2013, Nechaev2013,Michiardi2014}; the direct or indirect nature of the band gap~\cite{Aguilera2013_2, Nechaev2013}; orbital contributions~\cite{Kuo2021}; optical properties~\cite{ Plank2018,Mohelsky2024}; EELS~\cite{ Nechaev2015}; dispersion of the topological surface states~\cite{Aguilera2019}; and electron dynamics~\cite{ Battiato2017}. 

After the $GW$ calculations, we constructed tight-binding (TB) Hamiltonians whose parameters are obtained fully from first-principles with the help of Wannier functions. Wannier functions are linear combinations of the Bloch eigenfunctions and are defined in such a way that they are maximally spatially localized. To obtain the maximally localized Wannier functions (shown in Fig.~\ref{fig:WF}), we used the Wannier90 library~\cite{wannier90} embedded in the SPEX code. These Wannier functions form the basis in which the TB Hamiltonian is expressed. Wannier functions constitute a more natural and more accurate basis for TB Hamiltonians of solids than pure atomic orbitals. 

The construction of the TB parameters fully from first principles is done using the so-called Wannier interpolation technique. A formal description of the Wannier interpolation can be found in Ref.~\cite{Marzari2012}. The basic idea is as follows: The actual first-principles calculation of the DFT or $GW$ Hamiltonian is carried out on a coarse uniform $\mathbf{k}$~mesh. A set of selected bands is then transformed into maximally localized Wannier functions. The Hamiltonian is now expressed in real space in a basis of these Wannier functions. Thanks to the spatial localization of the Wannier functions, long-ranged interactions can be truncated allowing for an inverse Fourier transformation to
provide the Hamiltonian at an arbitrary $\mathbf{k}$~point. This is called the ``interpolated'' Hamiltonian and very often gives rise to physical and surprisingly accurate
band structures, which are practically indistinguishable from those obtained in an explicit first-principles calculation.

As we showed in Ref.~\cite{Aguilera2019}, these bulk Hamiltonians in real space can also be used to construct film Hamiltonians that provide exceptional accuracy in the description of bulk and surface states of this family of materials. 

\subsection{Calculation of Bloch Functions}
\begin{figure*}
        \centering
        \includegraphics{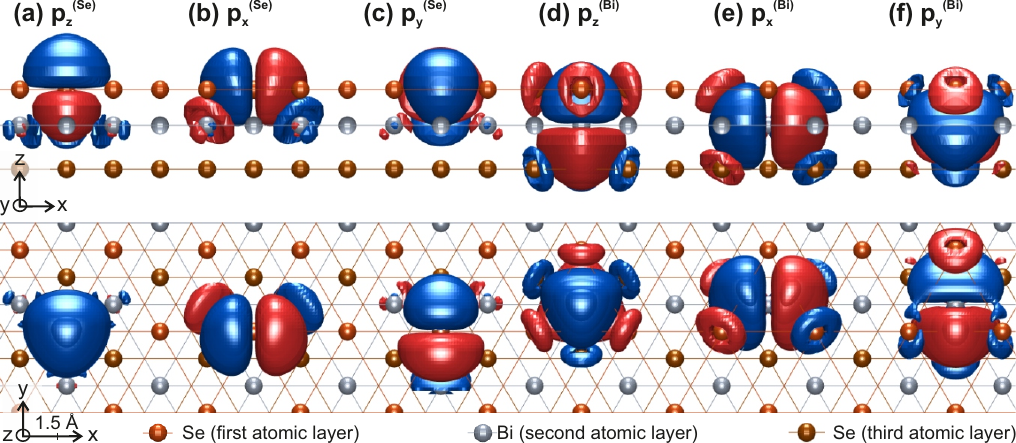}
        \caption{\label{fig:WF} Positive (red) and negative (blue) isosurfaces of the Wannier functions of \ce{Bi2Se3} localized on the Selenium and Bismuth atoms in the first and second atomic layers. 
        Orthographic projection onto $z$-$x$-plane in the top row and onto $y$-$x$-plane in the bottom row.
        The band structure close to $E_F$ is dominated by Wannier functions with $p_x$, $p_y$ and $p_z$ orbital character, such that the Wannier functions above are sufficient for the calculations in this work.
        The Wannier functions localized on the
        \ce{Se} atoms (light orange) extend further into the Van-der-Waals gap than those localized on the \ce{Bi} atoms (grey).
        } 
\end{figure*}

Mathematically, the interpolated TB Hamiltonian is a map that assigns to any $\bold{k}$ in the bulk Brillouin zone $\sf BZ$ a matrix acting on the electronic degrees of freedom. 
\begin{equation}
    \label{eq:H_bulk}
    \begin{split}
        \widehat{\rm H}\,:\, \sf{BZ} & \longrightarrow \mathrm{Mat}(30;\mathbb{C})\\
                 \bold{k} & \longmapsto \widehat{\rm H}(\bold{k}) = \sum_{\bold{R}\in \mathcal{L}} \rm{H}_\bold{R} \, e^{i \bold{k}\cdot \bold{R}}.
    \end{split}   
\end{equation}
Here $\mathcal{L}$ denotes the set of all lattice vectors and the hopping matrices $\rm{H}_{\bold{R}}$ are obtained fully from first principles as discussed in Section \ref{ssec:fpm} and given in the supplementary material of Ref.~\cite{Aguilera2019}. The basis functions, corresponding to the indices of the the hopping matrices $\rm{H}_{\bold{R}}$, are the following:
given by the $p$-orbitals $p^{i\sigma}_\alpha$ and ordered respectively according to the spin $\sigma =\, \uparrow,\downarrow$, to the atom number $i = 1, \ldots, 5$ on which the basis function is localized and to their spatial orientation $\alpha = z, x, y$. A selection of these orbitals is depicted in Fig.~\ref{fig:WF}.

The $s$-bands are energetically well separated from the $p$-bands, such that the orbitals given above are sufficient for a description of the band structure close to $E_F$~\cite{Aguilera2019}.
We note that the Hamiltonian in Equation \eqref{eq:H_bulk} is in the so called \emph{lattice gauge}.
The procedure to obtain $\rm H_\bold{R}$ can be performed after the DFT calculation or after the $GW$ correction.
When including many-body effects, the Hamiltonian, or more precisely the self-energy operator, is not necessarily self-adjoint, as it contains the lifetime of quasiparticle states.
In this work we only consider real eigenvalues of the many-body corrected TB Hamiltonians.

In the lattice gauge the Bloch functions can be written as a linear combination of the Wannier functions $w_\nu(\bold{r})$:
\begin{equation}
    \label{eq:BF_bulk}
    \Psi_{n,\bold{k}}(\bold{r}) = \sum_{\nu \in I} \mathcal{C}^n_\nu(\bold{k})\sum_{\bold{R}\in \mathcal{L}}  e^{i\bold{k}\cdot\bold{r}} w_\nu(\bold{r}-\bold{R}),
\end{equation}
where $\mathcal{C}^n(\bold{k})$ is the $n$-th eigenvector of the matrix $\widehat{\rm H}(\bold k)$. 
With Equation \eqref{eq:H_bulk} and \eqref{eq:BF_bulk} it is possible to analyze the the bulk electronic properties, for example as done in Figure~\ref{bulkBS} in Section~\ref{sec:mbc} of this article.
The Wannier functions $w_\nu(\bold{r})$ are obtained from first-principles 
as described in Section~\ref{ssec:fpm} and shown in Figure~\ref{fig:WF}.

To interpret the spectroscopic measurements presented in this article, the electronic structure of half-space or thick film systems is relevant. 
In this case the system size is so large, that it cannot be studied by $GW$ or DFT anymore.
With the hopping matrices $\rm H_\bold{R}$ it is possible to construct the TB Hamiltonian of samples with different geometries, by constructing the Hamiltonian as block matrix indexed by the lattice sites $\bold{R}_i$ and placing $\rm H_\bold{R}$ wherever $\bold R_i-\bold R_j = \bold R$.
In the case of a film system with a layered structure we maintain translational symmetry parallel to the sample surface.
Thus the Hamiltonian in this case is a block matrix populated close to the diagonal. The block entries are still dependent on the momentum $\bold k_\|$ parallel to the surface.
In this picture the on-diagonal terms correspond to intralayer terms, whereas the off-diagonal entries correspond to interlayer terms.
Alternatively, this can be viewed as performing an inverse Fourier transform of the Hamiltonian $\widehat{\rm H}$ of Equation~\eqref{eq:H_bulk} in the direction perpendicular to the surface and then restricting this operator with Dirichlet boundary conditions to a finite or semi-infinite $z$-range. 
Dirichlet boundary conditions correspond to a straightforward truncation of the matrix between the layer block matrices, leading to:
\begin{equation}
    \widehat{\rm H}_{z < 0}(\bold{k}_{\|}) = 
    \begin{pmatrix}
    \widehat{\rm H}_{11}(\bold{k}_{\|}) &\widehat{\rm H}_{12}(\bold{k}_{\|})& &  &\\
    \widehat{\rm H}_{21}(\bold{k}_{\|})& \widehat{\rm H}_{22}(\bold{k}_{\|})&\widehat{\rm H}_{23}(\bold{k}_{\|}) & &\\
     & \widehat{\rm H}_{32}(\bold{k}_{\|})& \widehat{\rm H}_{33}(\bold{k}_{\|})& &\\
     &  &   & \ddots
    \end{pmatrix},
\end{equation}
where the interlayer and intralayer terms are given analogous to Equation \eqref{eq:H_bulk}, but now only including vectors $\bold{R}: QL_i \rightarrow QL_j$, connecting unit cells from the $i$-th and $j$-th quintuple layers, in the sum.
\begin{equation}
    \widehat{\rm H}_{ij} = \sum_{\bold{R}: QL_i \rightarrow QL_j} \rm{H}_\bold{R} \,  e^{i \bold{k}\cdot \bold{R}}
\end{equation}
We note that, for materials in the \ce{Bi2Se3} family, which consist of weakly Van-der-Waals bound quintuple layers, Dirichlet boundary conditions are a natural and very precise approximation.
To keep our model as concise as possible, we neglect additional terms localized close to the boundary, such as a shift in the energies of the outermost orbitals.

The half-space Bloch functions are given \emph{inside} the bulk of the crystal as
\begin{equation}
    \Psi_{n,\bold{k}_\|}(\bold{r}) = \sum_{\nu \in I, \eta \in {1,..N_{QL}}} \mathcal{C}^n_{\nu,\eta}(\bold{k}_\|)\sum_{\bold{R}\in \mathcal{L}}  e^{i\bold{k}_\|\cdot\bold{r}} w_\nu(\bold{r}-\bold{R}),
\end{equation}
where $\mathcal{C}^n(\bold{k}_{\|})$ is the $n$-th eigenvector of the square matrix $\widehat{\rm H}_{z < 0}(\bold{k}_{\|})$ with size $30 \cdot N_{QL}$ and $w_\nu(\bold{r}-\bold{R})$ are the Wannier functions of the bulk material.
For the spectroscopic experiments a description of the Bloch functions \emph{outside} of a fictitious sample boundary is necessary. 
Calculating the Wannier functions of a film system or single quintuple layer from first-principles is computationally much more expensive than for the bulk.
Thus the propagation of the Bloch functions into the vacuum is not calculated self-consistently in this work.
We assume the laterally averaged potential $V$ at the boundary of the material to be: 
\begin{align}
    V(z) = \begin{cases}
        &\Phi   \quad \mathrm{for}  \quad z > z_{b}\\
        &0 \quad \mathrm{for}  \quad z < z_{b}
    \end{cases},
\end{align}
where $\Phi$ is the workfunction of the material and $z_b$ denotes the position of the fictitious sample boundary. 
We take for the workfunctions the values $\Phi_{BS} = \SI{5.6}{eV}$, $\Phi_{BTS} = \SI{5.25}{eV}$, $\Phi_{BT} = \SI{5.25}{eV}$ from Refs. \cite{Daichi2016, Ryu2018} and from the Gundlach oscillations reported in \cite{Gundlach1966, Binnig1985, Huang2013, SI}.  
We choose $z_b$ such that the boundary lies $\SI{1.3}{\angstrom}$ further out from the outermost atomic cores.
In the following we rely on the expression of Bloch functions as
\begin{equation}
\Psi_{n,\bold{k}_\|}(\bold{r}) = e^{i\bold{k}_\|\cdot\bold{r}} u_{n,\bold{k}_\|}(\bold{r}),
\end{equation}
where $u_{n,\bold{k}_\|}(\bold{r})$ has the periodicity of the lattice. Thus the function $u_{n,\bold{k}_\|}(\bold{r})$ restricted to the plane at the boundary of the sample is periodic, and can be developed into a two-dimensional Fourier series using the reciprocal surface lattice vectors $G_\Delta$. For $z$ \emph{outside} the fictitious sample boundary
\begin{equation}
\label{eq:FT}
\Psi_{n,\bold{k}_\|}(\bold{r}) = \sum_{\bold{G}_\Delta \in \mathcal{RL}} c^\Delta_{n,\bold{k}_\|} \,\, d^\Delta_{n,\bold{k}_\|}(z) \,\, e^{i(\bold{k}_\| + \bold{G}_\Delta) \cdot \bold{r}_\|}
\end{equation}
holds, with $d^\Delta_{n,\bold{k}_\|} (z) = \rm{exp}\left(-\kappa(n,\bold{k}_\|, \Delta) \, z\right)$ and the energy and momentum dependent decay constant given as 
\begin{equation}
    \label{eq:kappa}
    \kappa(n,\bold{k}_\|, \Delta) = \sqrt{\frac{2m}{\hbar^2}(E_{n,\bold{k}_\|}-E_F - \Phi) + (\bold{k}_\|+\bold{G}_\Delta)^2}.
\end{equation}

We calculate the Fourier coefficients $c^\Delta_{n,\bold{k}_\|}$ by numerically approximating the integral 
\begin{equation}
    c^\Delta_{n,\bold{k}_\|} = \int_{\sf{UC}} e^{-i \bold{G}_\Delta \cdot \bold{r}} u_{n,\bold{k}_\|}(x,y,z_b) \, \rm{dx} \, \rm{dy}
\end{equation}
over the two-dimensional surface unit cell $\sf{UC}$ with a finite sum over a $50 \times 50$ grid of $\bold{r}$ points in $\sf{UC}$ aligned with the surface lattice vectors.
The decay constant $\kappa(n,\bold{k}_\|, \Delta)$ can be deduced from the solution of the Schrödinger equation of a free particle \cite{Blügel1998}.
Upon examining Equation~\eqref{eq:kappa} it becomes apparent, that Bloch functions which lie far below the Fermi energy $E_F$ decay faster into the vacuum than those with higher energy. Also we see, that the Bloch functions with $\bold k$ close to $\sf \Gamma$ propagate further into the vacuum. Furthermore it is possible to truncate the sum in Equation~\eqref{eq:FT} after the first 3 reciprocal lattice vectors, as the plane wave components with larger $G_\Delta$ decay rapidly outside the sample, such that their contribution is negligible.

\subsection{Calculation of the Tunneling Current}
\label{ssec:cotc}
The starting point is Bardeen's formula for the tunneling current \cite{Bardeen1961, Chen1990}. The tunneling current $I$ is sensitive to all electronic states with energy $E$ between the Fermi energy $E_F$ of the sample and $E_F + eU_B$, where $U_B$ is the bias voltage applied to the tip sample junction.  
\begin{equation}
    \label{eq:I}
    I(U_B) = \mathrm{const.} \cdot \mathrm{sgn}(U_B) \cdot \sum_{E_\mu \in \mathcal{E}} \sum_{\nu \in \mathcal{I}} \vert M_{\mu \nu } \vert^2,
\end{equation}
where $\mathcal{E} = \big [\mathrm{min}(E_F,E_F + eU_B), \mathrm{max}(E_F,E_F + eU_B) \big ]$ and $\mathcal{I} = \{s,\, p_x,\, p_y,\, p_z\}$. For the differential conductance only the states close to $E_F + eU_B$ are relevant:
\begin{equation}
    \label{dIdV}
    \frac{dI}{dV}(U_B) = \mathrm{const.} \cdot \sum_{E_\mu \approx E_F + eU_B} \sum_{\nu \in \mathcal{I}} \vert M_{\mu \nu } \vert^2.
\end{equation}
The tunneling matrix element between the sample wavefunction $\psi_\mu$ and the tip wavefunction $\chi_\nu$ is given as an integral over a surface $\mathcal{S}$ separating tip and sample.
\begin{equation}
    \label{M}
    M_{\mu\nu} = -\frac{\hbar^2}{2m}\int_{\mathcal{S}}(\chi^*_\nu \nabla \psi_\mu -  \psi_\mu \nabla \chi^*_\nu)\,\bold{d}\mathcal{S}\, .
\end{equation}
Calculating this integral is computationally expensive and furthermore the precise tip wavefunctions $\chi_\nu$ are not known \cite{Mandi2015}. Chen's derivative rule is an efficient approximation, relying on the expansion of the tip wavefunction $\chi_\nu$ into spherical harmonics $Y_{lm}$ centered at the position of the tip apex $\bold{r}_0$. This expansion is given by the sum
\begin{equation}
    \label{SH}
   \chi_\nu(\bold{r})  = \sum_{ l m} C_{\nu l m}\,k_l(\kappa_\nu r)\, Y_{lm}(\theta,\varphi)\,
\end{equation}
where $r = \vert \bold{r}-\bold{r}_0\vert$, $\theta$ is the polar angle and $\varphi$ the azimuthal angle. The function $k_l(\kappa_\nu r)$ is the $l$-th Bessel function of the second kind, with $\kappa_\nu$ being the decay constant of the tip orbital into the vacuum. This expansion can also be done in terms of the real spherical harmonics with angular symmetries $\beta \in \{s, x, y, z, ...\}$, which are obtained as linear combinations of the complex-valued spherical harmonics. Introducing the notation $\widetilde{Y}_{\nu \beta} = k_\beta (k_\nu r) Y_{\beta}(\theta,\varphi)$ the sum in Equation \eqref{SH} is equivalent to 
\begin{equation}
    \label{RSH}
   \chi_\nu(\bold{r})  = \sum_{\beta } C_{\nu \beta}\,\widetilde{Y}_{\nu \beta}(\theta,\varphi)\,.
\end{equation}
Chens derivative rule gives an approximation for evaluating the integral in Equation~\eqref{M} \cite{Chen1990}. Based upon the expansion of the tip wavefunction $\chi_\nu(\bold{r})$ shown in Equation~\eqref{RSH} the tunneling matrix element can be approximated by
\begin{equation}
    \label{DR}
  \vert M_{\mu \nu}\vert^2 = \frac{4\pi^2\hbar^4}{\kappa_\nu^2 m_e} \bigg\vert \sum_\beta C_{\nu \beta} \, \hat{\partial}_{\nu \beta} \, \psi_\mu(\bold{r}_0)\bigg\vert^2\,.
\end{equation}
The differential operators $\hat{\partial}_{\nu \beta}$ act on the sample wavefunction $\psi_\mu$ and are evaluated at the position of the tip $\bold{r}_0$, they are given in Table \ref{tab:do}. As the Bloch functions $\Psi_{n,\bold{k}_\|}(\bold{r})$ are given as Fourier series in Equation~\eqref{eq:FT} the derivatives in $x$- and $y$-direction can be calculated by multiplying the Fourier coefficients $c^\Delta_{n,\bold{k}_\|}$ with the $x$- and $y$-components of $i\bold{k}_\| + \bold{G}_\Delta$. The $z$-derivative is obtained by multiplying the plane wave components $e^{i(\bold{k}_\| + G_\Delta)\cdot \bold{r}}$ of $\Psi_{n,\bold{k}_\|}(\bold{r})$ with $-\kappa(n,\bold{k}_\|, \Delta)$.

For clarification we note that the sum over the orbitals $\nu$ in Equation \eqref{eq:I} is a sum over magnitudes squared, whereas the sum in Equation \eqref{DR} over different spherical harmonics $\beta$ contributing to one orbital $\nu$ takes the relative phases into account.

\begin{table}
\vspace{3mm}
\centering
\renewcommand{\arraystretch}{1.5}
\renewcommand{\tabcolsep}{5mm}
\begin{tabular}{ |c|c|c|c|c|  }
 \hline
 \multicolumn{5}{|c|}{Differential Operators $\hat{\partial}_{\nu \beta}$ in the Derivative Rule} \\
 \hline
 $\beta$ & $s$ & $x'$ &$y'$& $z'$\\
 \hline
$\hat{\partial}_{\nu \beta}$  & $1$  & $\frac{1}{\kappa_\nu} \frac{\partial}{\partial x'}$&   $\frac{1}{\kappa_\nu} \frac{\partial}{\partial y'}$& $\frac{1}{\kappa_\nu} \frac{\partial}{\partial z'}$\\
 \hline
\end{tabular}
\caption{Correspondence between tip orbital character $\beta$ and differential operator $\hat{\partial}_{\nu \beta}$ according to Chens derivative rule. Prime coordinates correspond to the tip coordinate system.}
\label{tab:do}
\end{table}
For metal tips it is sufficient to consider only the tunneling current through the $s$-type orbital \cite{Chen1990}. 
In the case of \CO terminated tips the $p$-type orbitals also contribute significantly to the tunneling current \cite{Gross2011}. 
Orbitals with higher quantum number $n(\nu) \geq 2$ and corresponding differential operators with a higher order are not necessary for the tips used in the experiments presented here. 
The energy dependence of the constants $k_\nu$ and $C_{\nu \beta}$ is not known, we use $k_\nu = \SI{1}{\angstrom}$ and constant $C_{\nu \beta}$ as given in Table \ref{tab:EC}, similar to Ref.~\cite{Gross2011}. 
\begin{table}
\centering
\renewcommand{\arraystretch}{1.5}
\renewcommand{\tabcolsep}{5mm}
\begin{tabular}{ |c|c|c|c|c|  }
 \hline
 $\nu$ & $C_{\nu s}$ & $C_{\nu x'}$ &$C_{\nu y'}$& $C_{\nu z'}$\\
 \Xhline{0.6mm}
 \multicolumn{5}{|c|}{Expansion coefficients for metal tip} \\
 \hline
  $s$  & $1$  & $0$&   $0$& $0$\\
 \Xhline{0.6mm}
 \multicolumn{5}{|c|}{Expansion coefficients for \CO tip} \\
 \hline
  $s$  & $1/2$  & $0$&   $0$& $0$\\
 \hline
  $p_x$  & $0$  & $\sqrt{3/8}$&   $0$& $0$\\
 \hline
  $p_y$  & $0$  & $0$&   $\sqrt{3/8}$& $0$\\
 \hline
\end{tabular}
\caption{Expansion coefficients for the $s$-orbital of a metal tip and the $s$-, $p_x$- and $p_y$-orbital of a \CO tip in the tip coordinate system.}
\label{tab:EC}
\end{table}

In order to correctly describe the tunneling through \CO terminated tips the following considerations are necessary. The differential operators $\hat{\partial}_{\nu \beta}$ of Table~\ref{tab:do} are given in the tip coordinate system, which might be tilted with respect to the sample coordinate system. With prime coordinates $x', \, y', \, z'$ we denote the tip coordinate system, and with $x,\, y, \, z$ we denote the sample coordinate system. Following the conventions of Ref.~\cite{Mandi2015}, these coordinate systems are related to each other by
\begin{equation}
    \begin{pmatrix}
        x' & y'& z'
    \end{pmatrix}^{\mathrm{T}} = \mathrm{Rot}(\phi, \psi, \theta)  \begin{pmatrix}
        x & y & z
    \end{pmatrix}^{\mathrm{T}},
\end{equation}
where $\phi$, $\psi$ and $\theta$ are the Euler angles and the rotation matrix $\mathrm{Rot}(\phi, \psi, \theta)$ is defined in the supplemental material \cite{SI}.
There are two reasons which make it necessary to consider the tilt of the \CO tip. First, \CO tips are rarely perfectly cylindrically symmetric. Second, due to lateral forces acting on the apex atom the \ce{CO} molecule will bend by a varying amount as a function of the lateral tip position. This effect becomes significant for the tip-sample separations of $\SI{3}{\angstrom} - \SI{4}{\angstrom}$ which were used for the experiments described in this article \cite{Oinonen2024, Krejci2017}. According to Ref.~\cite{Mandi2015}, the expansion coefficients in the tip coordinate system $C_{\nu i'}$ are related to those in the sample coordinate system by
\begin{equation}
    \begin{pmatrix}
        C_{\nu x} \\
        C_{\nu y} \\
        C_{\nu z}
    \end{pmatrix} = \mathrm{Rot}(\phi, \psi, \theta)^{-1} \begin{pmatrix}
       C_{\nu x'} \\
        C_{\nu y'} \\
        C_{\nu z'}
    \end{pmatrix}.
\end{equation}

The relaxation of the \ce{CO} molecule upon approaching the sample can be calculated using the mechanistic probe-particle model (PPM) \cite{Hapala2014, Hapala2014_2}, which assumes that the interaction of the sample with the oxygen atom at the apex of the \CO tip is well described by a linear combination of Lennard-Jones potentials. In previous work we demonstrated that this model is suited very well for describing atomic forces that act between the tip and the samples under investigation in this work \cite{Liebig2022}. 
We note that extensions of the PPM are able to simulate STM images with a relaxation of the probe particle, however do not consider the accompanying tilt of the tip orbitals \cite{Krejci2017, Oinonen2024}.


\subsection{Experimental Methods}
\label{ssec:em}
The experiments were carried out on a commercial low-temperature combined scanning tunneling (STM) and atomic force microscope (AFM) (LT STM with qPlus option, Scienta Omicron GmbH, Taunusstein, Germany) which operates at a temperature of $\SI[locale=US]{4.4}{K}$ under ultra-high vacuum conditions at a base pressure of $1\cdot 10^{-11}\si{mbar}$. The microscope is equipped with a qPlus sensor for simultaneous STM and AFM measurements. The sensor used in this experiment (type qPlus M4) has a beam stiffness of $k = \SI{1850}{Nm^{-1}}$, as described in Table 1 of Ref.~\cite{Giessibl2019}. An electrochemically etched tungsten tip (wire diameter $\SI{50}{\micro m}$) is attached at the apex of the beam. The resonance frequency of the sensor is $f_0 = \SI[locale=US]{45.175}{kHz}$ with $Q \geq 350k$. The AFM was operated in the frequency modulation mode (FM-AFM) with a constant amplitude of $\SI{50}{pm}$. In the FM-AFM mode, for small oscillation amplitudes, the measured frequency shift $\Delta f$ of the sensor from its unperturbed resonance frequency $f_0$ is proportional to the gradient $k_{ts}$ of the vertical force $F_{ts}$ acting between tip and sample in the direction normal to the sample. The proportionality is given by $\Delta f = \frac{f_0}{(2k)}\langle k_{ts}\rangle$, whereby $\langle \,\, \rangle$ denotes a weighted average over the vertical oscillation of the tip.

For operating the STM, the bias voltage $U_B$ is applied to the sample. The differential tunneling conductance $(\mathsf{d}I/\mathsf{d}V)$ is measured by modulating the applied bias voltage by $U_{mod}$ at a frequency of $f_{mod}= \SI[locale=US]{5.464}{kHz}$, selected to avoid interference from noise peaks. The first harmonic of the tunneling current is detected by a lock-in amplifier (HF2-LI, Zurich Instruments, Zürich, Switzerland). The modulation amplitude is either $U_{mod} =\SI{10}{mV}$ or $U_{mod} =\SI{20}{mV}$, chosen as the best compromise between modulation broadening of the \dIdV signal and the signal-to-noise-ratio. The \dIdV signal is calibrated with the method described in \cite{Liebig2014, SI}.

The tips are prepared and characterized on a single crystal \ce{Cu} sample terminated in the (111) crystallographic plane, that was cleaned with standard sputter and anneal cycles. For tip characterization with the \ce{CO} front atom identification method (COFI) \cite{Hofmann2014} and tip functionalization a small amount of $\SI{0.01}{ML}$ of \ce{CO} was deposited on the \ce{Cu} surface. The tips were prepared with a series of pokes with a depth between $\SI{300}{\pico \m}$ and $\SI{10}{\nano \m}$ into the \ce{Cu}(111) surface as described in \cite{Liebig2020}. The number of \ce{Cu} atoms at the metal tip apex was determined with the COFI method. Before functionalizing the tip apex with a \ce{CO} molecule a tip terminating in a single copper atom was prepared. Subsequently the \ce{CO} molecule was attached to the tip with the method presented in \cite{Bartels1997}.
The electronic structure of the tips, in particular a constant density of states, was determined by referencing the \dIdV spectrum on the bare copper surface. 

After tip preparation and characterization, the \ce{Cu}(111) sample was exchanged with the topological insulator samples and the tip was approached in STM feedback mode. During this whole procedure the tip was kept cold to prevent any thermally-induced tip changes.

For the experiments on \ce{Bi2Se3} and \ce{Bi2Te2Se}  commercially available single crystal samples (HQGraphene, Groningen, Netherlands) with a purity $\geq 99.995\%$ were used. The samples were cleaved in the UHV chamber along the preferred (111) direction. Transfer to the low temperature microscope stage followed within minutes to prevent thermally activated diffusion and defect formation processes, as well as extrinsic adsorption of contaminants on the sample \cite{Zhu2011}. It has been demonstrated that this process yields large atomically flat surfaces with very few isolated defects \cite{Liebig2022}.

The \ce{Bi2Te3} sample was grown on a \ce{Al2O3}(0001) substrate via molecular beam epitaxy (MBE) similar to the growth process described in \cite{Mayer2021}. To protect the \ce{Bi2Te3} sample from oxidation and other unwanted adsorbates during the transport from the MBE to the UHV chamber of the STM/AFM, the surface is capped by a 10nm thick Tellurium layer. After installation the capping layer removal is done via thermal treatment at $\SI{270}{\degreeCelsius}$ for $\SI{45}{\min}$ in the preparation chamber of the STM/AFM. Subsequently the sample is cooled to room temperature and immediately transferred to the low temperature stage. This again yields large atomically flat surfaces free from unwanted adsorbates \cite{Grasser2021}.

\section{Scanning Tunneling Spectroscopy and Half-Space Band Structure}
\label{sec:STS}
\subsection{\ce{Bi2Se3}}
\label{Bi2Se3_STS}
\begin{figure}
        \centering
        \includegraphics{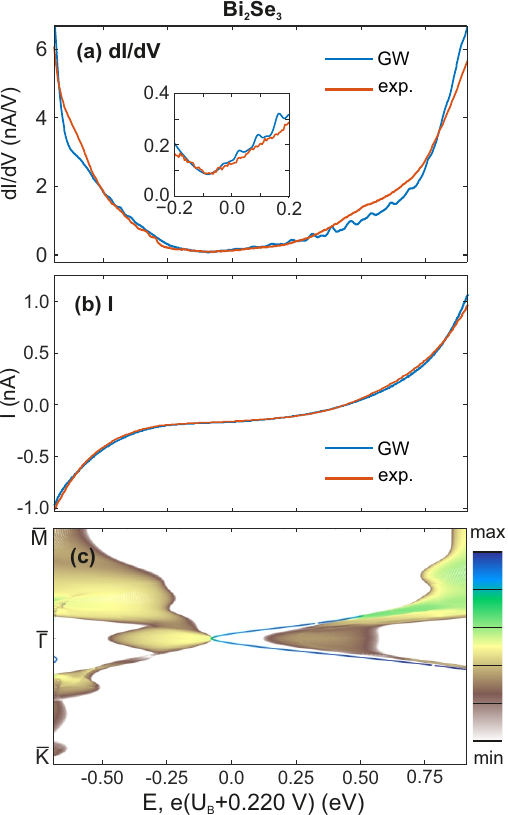}
        \caption{\label{Bi2Se3_dIdV} (a) \dIdV spectrum of a metal tip positioned above a Selenium atom of the \ce{Bi2Se3} surface overlayed  with the calculated spectrum. The inset shows the experimental and calculated $\mathsf{d}I/\mathsf{d}V$ for $\SI{-0.2}{V} < E < \SI{0.2}{V}$ with adjusted $y$-axis. (b) Experimental and calculated tunneling current. (c) Half-space band structure of \ce{Bi2Se3} color coded with $\mathrm{log}_{10}(\vert M_{\mu,\nu}(\mathbf{k}_{\|})\vert^2)$. Each black line in the color bar of (b) indicates one order of magnitude.}
\end{figure}

Unlike the other data and calculations presented in this article, Figure~\ref{Bi2Se3_dIdV} is concerned with a metal tip on the surface of \ce{Bi2Se3}.
This experiment is included to demonstrate that the methods used in this study are applicable to STS experiments with metal tips, which are commonly employed in preceding studies.

The calculated data in Figure~\ref{Bi2Se3_dIdV} (a) and (b) was obtained on a $90 \times 90$ Monkhorst-Pack grid. 
The colorcoding of the half-space band structure in Figure~\ref{Bi2Se3_dIdV} (c) indicates the base-$10$ logarithm of the square of the magnitude of the tunneling matrix element $M_{\mu, \nu}(\bold{k}_{\|})$, which forms the basis of the calculation.
Qualitatively, the \dIdV spectrum in Figure~\ref{Bi2Se3_dIdV}~(a) closely resembles the spectrum obtained using the \CO terminated tip shown in Figure~\ref{fig:intro}~(a). 
This similarity suggests that \CO terminated tips can in fact reliably reproduce \dIdV results otherwise measured with metal tips.

Figure~\ref{Bi2Se3_dIdV} (b) shows the tunneling current, obtained from the calculation $I^{(theo.)}(E)$ in blue and from experiment $I^{(exp.)}(U_B)$ in orange.
The theoretical current traces are calculated by integration the calculated $\mathsf{d}I/\mathsf{d}V$  from the experimental Fermi energy $E_F^{(exp.)}$ to the energy $E$.
These current traces are used to normalize the theoretical and experimental \dIdV spectra.
The method is described in detail in the supplemental material \cite{SI}.
A non-zero slope of $I(U_B)$ for all bias voltage $U_B$ indicates the closing of the band gap can be deduced from Figure~\ref{Bi2Se3_dIdV} (b), however the figure also demonstrates that the tunneling current is not adequate for an in-depth analysis of the electronic structure.

The theoretical and experimental \dIdV signal is much more suitable for this purpose. 
Figure~\ref{Bi2Se3_dIdV} (a) shows the  experimental spectrum of a metal tip on the surface of \ce{Bi2Se3} in orange and the calculated counterpart in blue. In experimental practice there is a shift of the Fermi level of the sample due to native point defects \cite{Liebig2022}. 
Thus to align the spectra we need to plot the energy $E$ and the shifted bias voltage $U_B + \Delta E_F$ on the abscissa of Figure~\ref{Bi2Se3_dIdV}. 
For the sample used during the measurements with a metal tip, shown in Figure~\ref{Bi2Se3_dIdV}, we find $\Delta E_F = \SI{0.22}{eV}$.
This is comparable to the sample used for the measurements in Figure~\ref{fig:intro} (a), where this shift was equal to $\SI{0.24}{eV}$.

Key features in the \dIdV spectrum are the minimum in the differential conductance at $E_D = \SI{-0.08}{eV}$ with a linear increase of the differential conductance away from $E_D$. 
Comparison with the half-space band structure in Figure~\ref{Bi2Se3_dIdV} (c) shows that this minimum in the \dIdV spectrum corresponds to the energy of the Dirac point $E_D$. 

Other important markers in the experimental \dIdV spectrum are at $E =  \SI{-0.7}{eV}$, $E =  \SI{-0.25}{eV}$ and $E = \SI{0.75}{eV}$, where the experimental \dIdV signal significantly increases towards lower and higher energy, respectively.
These features are reproduced by the theoretical \dIdV spectrum in Figure~\ref{Bi2Se3_dIdV} (a).
Comparison to the half-space band structures in Figure~\ref{Bi2Se3_dIdV} (c) shows that both increases are due to the bulk bands becoming less dispersive in the $\overline{\mathsf{\Gamma}}$--$\overline{\mathsf{M}}$ direction.
This flattening of the bands leads to accumulation of states on the energy axis and thus to an increased differential conductance.

However, we also note that there are discrepancies between the calculated and experimental spectra visible in Figure~\ref{Bi2Se3_dIdV}~(a).
The major discrepancy is, that the calculated \dIdV spectrum is significantly lower than the experimental spectrum for $E = 0.5 \pm 0.1  \si{eV}$.
We first explain the behaviour of the theoretical spectrum in this energy range and then focus on possible explanations of this discrepancy.
In the theoretical calculation of Figure~\ref{Bi2Se3_dIdV} (a) we see an approximately linear increase in the \dIdV signal for $\SI{-0.09}{eV} < E < \SI{0.75}{eV}$.
In the $\overline{\mathsf{\Gamma}}$--$\overline{\mathsf{K}}$ direction the topological boundary mode remains distinct from the bulk modes up to $E = \SI{0.75}{eV}$ and the tunneling matrix element associated to it retains a large magnitude.
In the $\overline{\mathsf{\Gamma}}$--$\overline{\mathsf{M}}$ direction the topological boundary mode hybridizes with the bulk modes at $E = \SI{0.49}{eV}$.
Following the linear dispersion relation of the topological boundary mode there is a set of bulk conduction bands associated to an increased tunneling probability, which appears in green in the color scale of Figure~\ref{Bi2Se3_dIdV} (c).
This is far beyond the conduction band minimum, which is located at $ E = \SI{0.15}{eV}$.
Our calculation of the tunneling matrix elements associated to the conduction band edge shows a very small magnitude, such that states near the conduction band edge do not contribute to the calculated \dIdV spectrum.
The dispersion relation of the topological boundary mode in these two high-symmetry directions and the vanishing of the tunneling matrix element close to the CBM explain why the increase of the  differential conductance follows the linear trend up to $\SI{0.75}{eV}$.
However, the maximum of the deviation at $E = 0.5\pm 0.1 \si{eV}$ coincides with the presence of many flat conduction bands in the vicinity of $\sf \Gamma$.
A slight deviation in the calculation of the tunneling matrix element of these bands, for example due to surface effects, might thus nonetheless explain the discrepancy between the experimental and calculated spectrum.

Another, rather minor discrepancy is the presence of small peaks in the calculated spectrum.
These are numerical artefacts caused by the finite grid of $\bold{k}$ points used to approximate the Brillouin zone integral.

\begin{figure}
        \centering
        \includegraphics{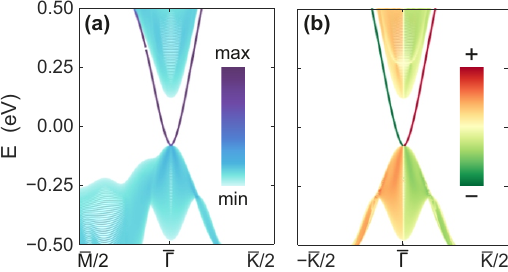}
        \caption{\label{fig:Bi2Se3_DC} (a) Half-space band structure of \ce{Bi2Se3} with color map representing the localization on the first QL. Following the linear dispersion of the upper Dirac cone there are darker lines inside the valence bands for $\SI{-0.08}{eV} > E > \SI{-0.25}{eV}$ which represent the strongly hybridized surface states. (b) Band structure from $-\overline{\mathsf{K}}/2$
        to $\overline{\mathsf{K}}/2$ with color indicating expectation value on spin orthogonal to $\bold{k}$. The spin polarization in the lower Dirac cone is minimal due to the hybridization with the bulk bands.}
\end{figure}

Finally, we want to discuss the topological boundary mode and its contribution to the \dIdV signal in more detail.
Both of the linear sections around $E_D$, clearly visible in the inset of in Figure \ref{Bi2Se3_dIdV} (a),  are well reproduced by the calculation shown in blue. 
As discussed for the spectrum of a \CO terminated tip in Section \ref{sec:intro}, we note that our calculation shows that the linear section for $E > E_D$ corresponds to tunneling into and out of the topological boundary mode. 
This interpretation agrees with previous reports \cite{Dmitriev2014}. 
For energies $\SI{-0.08}{eV} > E > \SI{-0.25}{eV}$ the topological surface state is strongly hybridized with parabolic bulk valence bands. 
In Figure~\ref{fig:Bi2Se3_DC} (a) the topological boundary mode appears as dark purple line crossing the bulk band gap with a linear dispersion. The valence band edge appears in light blue. 
The $GW$ calculation indicates the Dirac point is located at the VBM.
Below the VBM we see two darker features inside the valence band edge that continue along the linear dispersion of the topological boundary mode for $\SI{-0.08}{eV} > E \SI{-0.25}{eV}$. These dark features correspond to the hybridization of the Dirac cone with the bulk bands. 
The color map of Figure~\ref{fig:Bi2Se3_DC} (b) indicates the expectation value on the spin component orthogonal to $\bold k$. 
In accordance with the previous observation the spin momentum polarization of the Dirac cone decreases rapidly upon hybridization with the bulk bands. 
The calculation of the theoretical \dIdV spectrum shows that tunneling out of these hybridized bands reproduces the linear increase for $E < E_D$ in the \dIdV spectrum. 
This is a modification of previous reports, were the lower linear section was attributed to a possible isolated lower half of the Dirac cone \cite{Dmitriev2014}. 

Summarizing these results, the dispersion of the topological boundary mode close to the Dirac point can be studied using STS and the observations are consistent with $GW$ calculations of the band structure. Additionally, the \dIdV spectrum reveals the energies where the dispersion of the bulk conduction and valence bands weakens, consistent with the calculated half-space band structure.






\subsection{\ce{Bi2Te2Se}}
\label{Bi2Te2Se_STS}

\begin{figure}[h]
        \centering
        \includegraphics{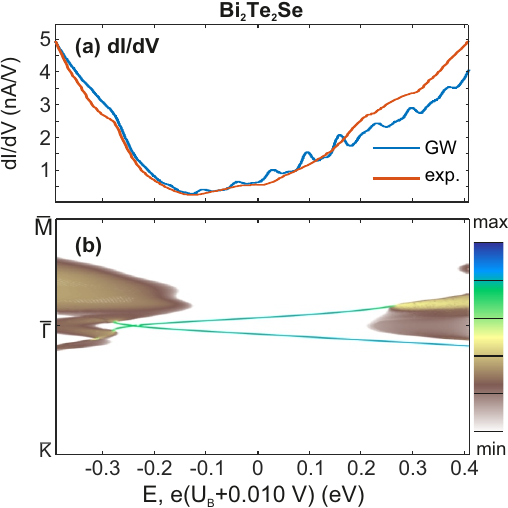}
        \caption{\label{fig:dIdV_Bi2Te2Se} (a) \dIdV spectrum of a \CO tip positioned above a \ce{Te} atom in the first atomic layer sites on the \ce{Bi2Te2Se} surface. (b) half-space band structure of \ce{Bi2Te2Se} around the bulk band gap color coded with $\mathrm{log}_{10}(\vert M_{\mu,\nu}(\mathbf{k}_{\|})\vert^2)$. Each line in the color bar of (b) indicates one order of magnitude.}
\end{figure}

Figure \ref{fig:dIdV_Bi2Te2Se} (a) shows the experimental and calculated \dIdV spectrum of a \CO terminated tip positioned on a \ce{Te} atom on the surface of \ce{Bi2Te2Se}. 
An atomic site with maximal distance to substitutional defects was chosen for this. The spectrum was recorded on a bias range from $\SI{-0.4}{V}$ to $\SI{0.4}{V}$. 
The corresponding half-space band structure is displayed in Figure~\ref{fig:dIdV_Bi2Te2Se} (b) with colorcoding analogous to that in Subsection~\ref{Bi2Se3_STS}.
The topological boundary mode of \ce{Bi2Te2Se} appears in blue and green, which indicates the strong contribution of this mode to the differential conductance.  The good agreement between theoretical and experimental characterization of the $\sf LDOS$ leads us to the following conclusions. 

From the comparison in Figure~\ref{fig:dIdV_Bi2Te2Se} a dopant induced shift of the Fermi level by $\Delta E_F = \SI{0.01}{eV}$ can be determined, indicating a low level of bulk doping.

The energy of minimal differential conductance is $\SI{-0.12}{eV}$.
Comparison to the half-space band structure conclusively shows that this does not  correspond to the Dirac point. 
For energies below $\SI{-0.12}{eV}$ tunneling into the bulk valence bands leads to an increase in the differential conductance. 
The position of the Dirac point is $E_D = \SI{-0.22}{eV}$, deep inside the bulk valence bands. 
At the Dirac point there is a plateau or shoulder like feature in the experimental and calculated \dIdV spectrum. 
We attribute this to the flattening in the dispersion relation as the boundary mode hybridizes with the bulk modes. 
Also at this energy there are weakly dispersing surface resonances of the bulk valence bands visible in the $\overline{\mathsf{\Gamma}}$--$\overline{\mathsf{K}}$ direction, which could also be a cause for observing this plateau-like feature.

As is the case for \ce{Bi2Se3}, the tunneling matrix element of states at the bulk conduction band edge have a negligible contribution to the \dIdV signal.
The behaviour of the topological boundary mode relative to the bulk conduction bands is similar to \ce{Bi2Se3} as well.
In the $\overline{\mathsf{\Gamma}}$--$\overline{\mathsf{K}}$ direction the boundary mode remains distinct from the bulk modes within the investigated energy range.
Whereas in the $\overline{\mathsf{\Gamma}}$--$\overline{\mathsf{M}}$ direction the boundary mode resonates strongly with the bulk modes, which can be seen as yellow region in bulk conduction bands, which continues along the linear dispersion of the boundary mode.
As a result, for $\SI{-0.12}{eV} < E < \SI{0.4}{eV} $ the experimental and calculated \dIdV spectra follow a linear trend.
Above $E = \SI{0.2}{eV}$ there are some deviations from the linear trend in the experimental \dIdV signal, which are not reproduced by the calculated spectrum.
Possible reasons for this deviation include tunneling into bulk modes, that is not entirely described by our model, or the large density of substitutional defects in our \ce{Bi2Te2Se} sample.
Also it was reported that a (trivial) two-dimensional electron gas (2DEG) can emerge at the surface of \ce{Bi2Se3}, a compound very similar to \ce{Bi2Te2Se}, due to band bending at the interface to the vacuum surrounding the sample \cite{Bianchi2010}. Energetically this 2DEG occurs at the conduction band edge, which is where we observe the deviation of experiment and theory in Figure~\ref{Bi2Te2Se_STS} (a), suggesting that it may explain the discrepancy.
Based on the current analysis, it remains unclear which of these mechanisms is most relevant.

The energy range of Figure~\ref{fig:dIdV_Bi2Te2Se} allows for a detailed investigation of the boundary mode and the Dirac cone, however it is not large enough for observing the flattening in the dispersion of the bulk bands, as we discussed previously in Subsection~\ref{Bi2Se3_STS}. 

\subsection{\ce{Bi2Te3}}
\label{Bi2Te3_STS}

\begin{figure}
        \centering
        \includegraphics{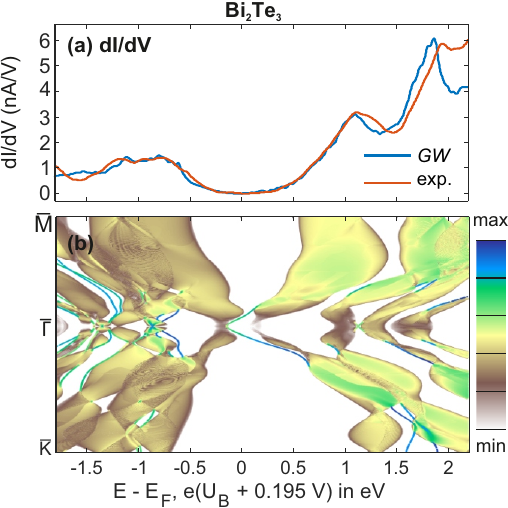}
        \caption{\label{Bi2Te3_dIdV} (a) \dIdV spectrum of a \CO tip positioned above a hollow site on the \ce{Bi2Te3} surface. (b) half-space band structure of \ce{Bi2Te3} color coded with $\mathrm{log}_{10}(\vert M_{\mu,\nu}(\mathbf{k}_{\|})\vert^2)$. Each black line in the color bar of (b) indicates one order of magnitude.}
\end{figure}

Figure \ref{Bi2Te3_dIdV} (a) shows the \dIdV spectrum of \CO terminated tip positioned above an hollow site on the \ce{Bi2Te3} surface over a bias voltage range of $\SI{-2}{V}$ to $\SI{2}{V}$. The experimental spectrum exhibits peaks in the differential conductance at $\SI{-1.2}{V}$, $\SI{-0.9}{V}$, $\SI{1.1}{V}$ and $\SI{1.75}{V}$. 
The orbital character of these peaks is discussed in Section~\ref{ssec:specm}.

Figure~\ref{Bi2Te3_dIdV}~(b) shows the $GW$ band structure of \ce{Bi2Te3}. 
The topological boundary mode appears in blue and green in the color coding of Figure~\ref{Bi2Te3_dIdV}~(b).
The bulk valence bands appear in yellow and brown.
In contrast, the bulk conduction bands are shown in mostly green, which corresponds to a larger magnitude of $M_{\mu,\nu}(\mathbf{k}_{\|})$.
This can be explained by a slower decay of the wave functions $\Psi_{n,\bold{k}}$ into the vacuum barrier, if $E_n(\bold{k})$ is larger. 
This relation is given quantitatively by Equation~\eqref{eq:kappa} in Section~\ref{ssec:cotc}, and thus can be understood on the level of effective non-interacting particles. 
As the $\sf{DOS}$ is approximately the same below and above $E_F$, this leads to a general trend in the \dIdV spectrum of increasing differential conductance towards more positive energies in Figure~\ref{Bi2Te3_dIdV}~(a).

We associate the peaks in the \dIdV spectrum with the accumulation of bulk bands and the flattening of their dispersion near the Brillouin zone boundary. 
For $E > 0$, there is a gap of approximately $\SI{0.5}{eV}$ between the lowest unoccupied bands and the next group of unoccupied bands as they approach the Brillouin zone boundary. 
This gap leads to the emergence of two distinct peaks at positive energies in the \dIdV spectrum shown in Figure~\ref{Bi2Te3_dIdV} (a), separated by roughly the same energy gap. 
In contrast, for $E < 0$, the bulk bands exhibit significant overlap, resulting in less distinct peaks at negative energies. 
Additionally, we observe a Rashba and other surface states for $E < \SI{-0.5}{eV}$ in the vicinity of the $\sf \Gamma$ point as blue and green lines in Figure~\ref{Bi2Te3_dIdV} (b).
We assume that the contribution of these bands is small compared to the bulk bands.
An experimental justification for this assumption can be derived from comparing the \dIdV signals across energy ranges where tunneling predominantly occurs into the topological boundary mode with respect to tunneling into bulk modes. 
The signal from tunneling into the bulk modes outweighs that from the topological boundary mode, despite the larger tunneling matrix element associated with the topological boundary mode. 
The magnitude of the tunneling matrix element for the non-topological surface states is of similar magnitude as the tunneling matrix element of the topological boundary mode. 
Therefore, it is reasonable to conclude that tunneling into the bulk modes is more significant in determining the overall \dIdV signal.

To the best of our knowledge only very few experimental \dIdV measurements on this material class over such a large bias range have been published so far.
There are experimental difficulties in obtaining spectra over this bias range. Due to the high bias voltage applied to the tip sample junction, large electrostatic forces act and a high current of up to $\SI{2}{nA}$ flows. 
Both effects increase the risk of a tip change or loss of the attached \ce{CO} molecule. 
In our experiments we increased the tip sample separation by $\SI{120}{pm}$ to limit the magnitude of the tunneling current and electrostatic forces.
Furthermore, the likelihood of a spurious electronic tip state distorting the \dIdV spectrum increases with increasing energy range. 

Despite this, studying spectra across extended energy ranges provides valuable insights. 
For instance, Ref.~\cite{Netsou2020} examines energy level shifts due to varying stoichiometry. 
Their theoretical density of states calculations span a similar energy range as shown in Figure~\ref{Bi2Te3_dIdV}, whereas the experimental data of Ref.~\cite{Netsou2020} covers only a much smaller range.
Focusing only on a narrow energy scale is problematic since \dIdV spectra near the bulk band gap are generally featureless. 
Consequently, shifts in a single feature or changes in the vacuum barrier could be misinterpreted as shifts of the entire spectrum.

\section{Probing Many-Body Corrections}
\label{sec:mbc}

Our analysis in Section~\ref{Bi2Se3_STS} shows that drastic changes in the half-space band structure of \ce{Bi2Se3} close to the Dirac point due to the $GW$ correction are in agreement with experimental observations.
However, this does not provide the means to quantitatively show that $GW$ aligns better with experiment than DFT.
For a quantitative comparison we use the observation described in the previous section \ref{Bi2Te3_STS} at the example of \ce{Bi2Te3}. 
Groups of bulk bands can be associated with peaks in the \dIdV spectrum and thus the position of these peaks can be used as a measure for energy gaps in the band structure.

\begin{figure}
        \centering
        \includegraphics{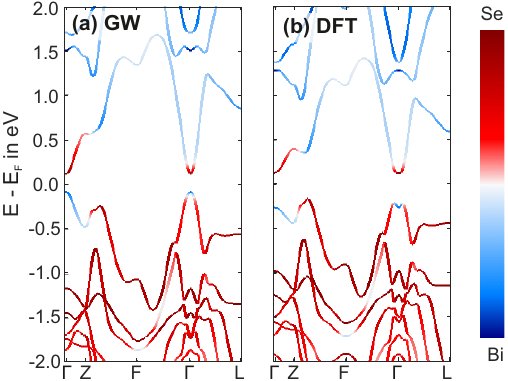}
        \caption{\label{bulkBS} (a) $GW$ band structure and (b) DFT-band structure  of bulk \ce{Bi2Se3}. DFT tends to underestimate trivial band gaps and overestimates non-trivial band gaps. The $GW$ correction thus decreases the band gap around $\sf \Gamma$ but increases the separation between conduction and valence bands in the rest of the Brillouin zone. Furthermore the $GW$ correction removes the camelback shaped dispersion at the $\sf \Gamma$-point.}
\end{figure}

First, we describe the main effects of the $GW$ correction on the bulk band structure.
In Figure~\ref{bulkBS} both the bulk $GW$ band structure (a) and  the bulk DFT band structure (b) is plotted. 
The color code represents the orbital character of the Bloch functions.
Blue indicates the corresponding Bloch function is primarily a linear combination of Wannier functions localized on \ce{Bi} atoms and red indicates localization on the \ce{Se} atoms.
As expected, the valence bands primarily arise from the $3p$ orbitals of \ce{Se}, and the conduction bands predominantly involve the $5p$ orbitals of \ce{Bi}. 
In both band structures the band inversion at the $\sf \Gamma$-point is evident.
The $GW$ correction tends to increase the magnitude of non-inverted energy gaps while decreasing the magnitude of inverted gaps \cite{Aguilera2019}.
In topological insulators, this correction significantly alters the dispersion relation, as both inverted and non-inverted energy gaps are present in the band structure.

The tendency of $GW$ to increase the magnitude of non-inverted energy gaps can be determined from the position of the bulk conduction bands at the $\sf L$- and  $\sf F$-point. 
According to the DFT calculation the energy of the lowest unoccupied band at the $\sf L$-point is $E = \SI{-0.43}{eV}$ and of the highest occupied band is $E = \SI{0.60}{eV}$, which leads to an energy gap of $\Delta E^{\mathsf{L}}_{\mathrm{DFT}} = \SI{1.03}{eV}$
In the $GW$ band structure the energy of the lowest unoccupied band at the $\sf L$-point is $E = \SI{0.85}{eV}$ and of the highest occupied band is $E = \SI{-0.55}{eV}$. Thus the energy gap of in the $GW$ band structure is $\Delta E^{\mathsf{L}}_{GW} = \SI{1.4}{eV}$, which is larger than $\Delta E^{\mathsf{L}}_{DFT} = \SI{1.03}{eV}$ by about $36\%$.
At the $\sf F$ point we have $\Delta E^{\mathsf{F}}_{GW} = \SI{2.4}{eV}$ and $\Delta E^{\mathsf{F}}_{\mathrm{DFT}} = \SI{2.0}{eV}$.

The influence of the $GW$ correction on the value of the inverted band gap $g$ at the $\sf \Gamma$ point is reversed. 
We find $g_{GW} = \SI{0.21}{eV}$ and $g_{\mathrm{DFT}} = \SI{0.24}{eV}$
Furthermore the dispersion relation of the bulk conduction band at the $\sf \Gamma$-point is no longer camelback shaped in the $GW$ band structure.

\begin{figure}
        \centering
        \includegraphics{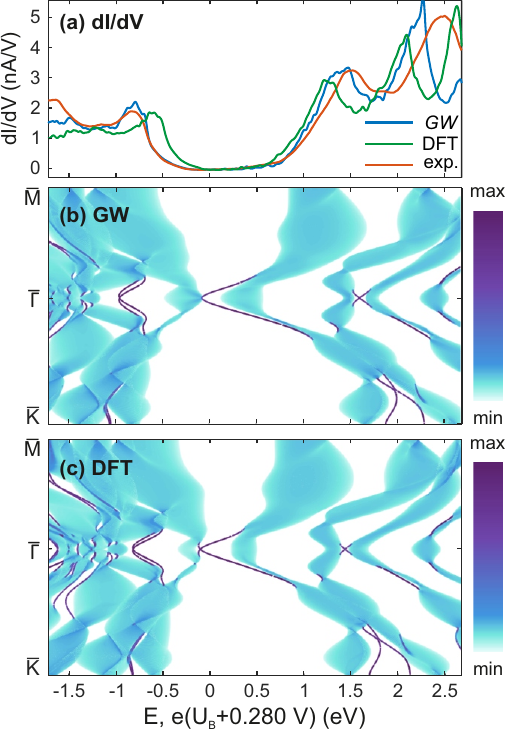}
        \caption{\label{GWvsDFT_large} (a) Comparison of experimentally obtained \dIdV spectrum of a \CO tip positioned above a \ce{Se} atom in the first atomic layer (orange) with the theoretical spectra calculated from the $GW$ Hamiltonian (blue) and DFT-Hamiltonian (red). (b) half-space band structure of the $GW$ Hamiltonian for \ce{Bi2Se3}. (c) half-space band structure of the DFT-Hamiltonian for \ce{Bi2Se3}.}
\end{figure}
The observations made for the bulk band structure also apply to the half-space band structures shown in Figure~\ref{GWvsDFT_large} (b) and (c). 
Note that here the lowest unoccupied and the highest occupied band smear out, but especially the lowest unoccupied bands are recognizable as distinct group of bands, separated from the rest of the band structure.
The energy gap between the bulk modes, shown in light blue, at $\overline{\sf K}$ and $\overline{\sf M}$ is smaller in the DFT-band structure than in the $GW$ band structure. 
More precisely we have $\Delta E^{\overline{\mathsf{K}}}_{GW} = \SI{2.4}{eV}$ and $\Delta E^{\overline{\mathsf{M}}}_{GW} = \SI{1.41}{eV}$ in the $GW$ band structure. 
In the DFT band structure we find $\Delta E^{\overline{\mathsf{K}}}_{\mathrm{DFT}} = \SI{2.03}{eV}$ and $\Delta E^{\overline{\mathsf{M}}}_{\mathrm{DFT}} = \SI{1.07}{eV}$.
Furthermore the M-shaped dispersion relation at the $\sf \Gamma$-point in the case of the bulk DFT band structure is retained in the half-space DFT band structure. 
As a consequence of the inverted band curvature at $\sf \Gamma$ in the half-space DFT band structure, the Dirac point is buried beneath the valence band maximum. 

The calculated and experimental \dIdV spectra are plotted in Figure~\ref{GWvsDFT_large} (a). 
In the experimental spectrum, shown as orange curve, we observe local maxima in the differential conductance at $\SI{-0.8}{eV}$, $\SI{1.5}{eV}$ and $\SI{2.4}{eV}$. 
We denote by $\Delta E_{exp} = \SI{2.3}{eV}$ the difference between the peak at negative energies and the first peak at positive energies.
We note that it is possible to distinguish \ce{Bi2Se3} from \ce{Bi2Te3} on the basis of the \dIdV spectra in Figure~\ref{Bi2Te3_dIdV} and Figure~\ref{GWvsDFT_large}, as these three peaks are much farther apart in the case of \ce{Bi2Se3}. 
The \dIdV spectrum, which was calculated using the $GW$ Hamiltonian, is plotted in blue in Figure~\ref{GWvsDFT_large} (a), whereas the DFT calculation is the green line in Figure~\ref{GWvsDFT_large} (a).

The $GW$ calculation exhibits a closer alignment with the experimental data.
$GW$ reproduces the maxima in the experimental \dIdV spectra at $E = \SI{-0.8}{eV}$ and $E = \SI{1.5}{eV}$ with the correct position and amplitude.
Furthermore $GW$ follows the development of the experimental \dIdV signal around the bulk band gap closely.
There is a sharp increase visible in the experimental \dIdV spectrum at \SI{0.75}{eV} which is reproduced by the $GW$ calculated \dIdV spectrum.
As established in Section~\ref{Bi2Se3_STS} this feature is due to the flattening of the dispersion of the bulk conduction bands in the $\overline{\sf{\Gamma}}$ -- $\overline{\sf{M}}$ direction.
However, the $GW$ calculation predicts the third peak in the spectrum to be at $E = \SI{2.25}{eV}$, whereas this peak is located at $E = \SI{2.5}{eV}$ in the experimental spectrum.

In contrast, the DFT calculation shows a noticeable shift in energy for all the observed peaks.
Furthermore the trench
in the DFT \dIdV spectrum around the bulk band gap is narrower than in the experimental spectrum, reflecting the tendency of DFT to underestimate trivial gaps.

The difference of the two maxima around the bulk band gap in the experimental \dIdV spectrum is $\Delta E_{exp} = \SI{2.3}{eV}$. 
We find that the numerical value of $\Delta E_{exp}$ lies in between the band gaps of the $GW$ band structure at the $\overline{\mathsf{K}}$ and $\overline{\mathsf{M}}$ points, \emph{i.e.},
\begin{equation}
    \Delta E^{\overline{\mathsf{K}}}_{GW} < \Delta E_{exp} < \Delta E^{\overline{\mathsf{M}}}_{GW}.
\end{equation}
In contrast we find, that both of the gaps in the DFT band structure,  $\Delta E^{\overline{\mathsf{K}}}_{\mathrm{DFT}}$ and $\Delta E^{\overline{\mathsf{M}}}_{\mathrm{DFT}}$, are smaller than $\Delta E_{exp}$. 
We conclude that DFT is able to qualitatively reproduce the shape of the experimental \dIdV spectrum, but fails when it comes to predicting the quantitative energy gaps.

\section{Orbital Composition}
\label{sec:orbital}
Due to the very high spatial resolution of AFM and STM it is possible to apply the previously presented results to determine the orbital contributions to the band structure. 
In the \emph{spectral} method presented in Section~\ref{ssec:specm} we compare the \dIdV signal as a function of energy with the tip being positioned above different inequivalent lattice sites. 
In Section~\ref{ssec:spatm} the \emph{spatial} method is introduced, where the \dIdV signal is obtained as a function of the lateral tip position and we compare maps at different bias voltages $U_B$. 

An experimental approach to achieve similar experimental results using ARPES is presented in Ref.~\cite{Kuo2021}. 
In standing-wave excited hard X-ray photoemission spectroscopy (SW-HAXPES), atoms in the bulk crystal are excited using standing waves. Individual layers are excited with different intensity, due to the distribution of nodes and antinodes of the standing wave pattern across the atomic layers. 
Due to the layered structure of \ce{Bi2Se3}, the band structure can be studied element resolved using the technique described in Ref.~\cite{Kuo2021}.
However, opposed to the experimental methods presented here, \mbox{SW-HAXPES} cannot be used to probe the unoccupied conduction bands.

\subsection{Spectral Method}
\begin{figure}
    \centering
    \includegraphics{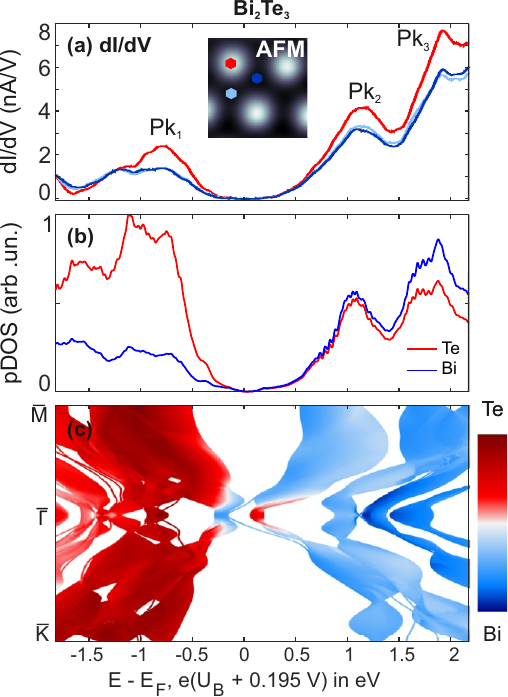}
    \caption{\label{fig:oc_spectral} (a) \dIdV spectra of a \CO terminated tip on the three nonequivalent crystal sites on the \ce{Bi2Te3} surface. (b) Partial Density of States corresponding to the Wannier functions localized on the \ce{Te} and \ce{Bi} positions in the first and second atomic layers. (c) half-space band structure of \ce{Bi2Te3} with color representing orbital character of the energy bands.}
\end{figure}

Naturally, STS is not confined to studying only the occupied electron states, as electrons can tunnel from the tip into the sample and vice versa.
Thus the method we present here can be applied to characterize the orbital structure of the valence band relative to the conduction band.
The experimental data underlying this analysis consist of \dIdV spectra obtained on a \ce{Te} site (red) and the two nonequivalent off-sites O1 and O2 (shades of blue), as shown in Figure~\ref{fig:oc_spectral} (a).
The inset shows the frequency shift channel of an AFM measurement, where the \ce{Te} atoms are imaged as bright spots.
The lateral positions at which the spectra have been measured are indicated with the the corresponding colors.

\begin{table}
    \centering
    \renewcommand{\arraystretch}{1.5}
    \renewcommand{\tabcolsep}{5mm}
        \begin{tabular}{ |c|c|c|c|c|  }
         \hline
             & $\mathsf{Pk_1}$ & $\mathsf{Pk_2}$ & $\mathsf{Pk_3}$ \\
         \hline
        \ce{Te}  & $\SI{2.39}{nA/V}$  & $\SI{4.14}{nA/V}$&   $\SI{7.71}{nA/V}$\\
         \hline
         O1  & $\SI{1.41}{nA/V}$  & $\SI{3.32}{nA/V}$&   $\SI{5.64}{nA/V}$\\
         \hline
         O2  & $\SI{1.41}{nA/V}$  & $\SI{3.17}{nA/V}$&   $\SI{5.90}{nA/V}$\\
         \hline
          $\mathcal{R}$  & $59\%$  & $(77 \text{--} 80)\%$ &   $(73 \text{--} 77)\%$\\
         \hline
        \end{tabular}
    \caption{Differential conductance ate the peaks of the spectra shown in Figure~\ref{fig:oc_spectral} (a) and the ratio $\mathcal{R}$ of the peak height at the hollow sites O1 and O2 relative to the \ce{Te} atomic site.}
    \label{tab:ph}
\end{table}

The characteristic peaks, marked by $\mathsf{Pk_1}$, $\mathsf{Pk_2}$ and $\mathsf{Pk_3}$, as discussed in Section~\ref{Bi2Te3_STS} and Section~\ref{GWvsDFT_large}, can be attributed to groups of weakly dispersing bulk bands. Table~\ref{tab:ph} lists the maximum differential conductance at these peaks for each crystal site.
Furthermore, for a direct comparison the ratio 
\begin{equation}
    \mathcal{R} = \frac{\mathsf{d}I/\mathsf{d}V(\mathrm{O}1,\mathrm{O}2)}{\mathsf{d}I/\mathsf{d}V(\ce{Te})}
\end{equation}
is given. 
The range of values for $\mathsf{Pk_2}$ and $\mathsf{Pk_3}$ results from the slight difference in the \dIdV signal above O1 and O2.

For the first peak $\mathsf{Pk_1}$ at negative energy we see a drastic reduction of the peak amplitude giving $\mathcal{R} = 59\%$, when moving the tip from the \ce{Te} site to the offsite.
This can be explained by the fact that the valence bands are formed by primarily \ce{Te} orbitals.
This can be seen from the partial $\sf DOS$ shown in Figure~\ref{fig:oc_spectral} (b), which indicates that the partial $\sf DOS$ associated to the \ce{Te} degrees of freedom is much larger than the partial $\sf DOS$ associated to \ce{Bi} for negative energies $ E < \SI{-0.5}{eV}$.

\label{ssec:specm}

However, also for the peaks at positive energies we observe a reduction in the differential conductance when moving away from the \ce{Te} site.
Due to the exponential localization of the Wannier functions, the main contribution to \dIdV signal comes from the components of the Bloch functions localized on the \ce{Te} atoms in the first atomic layer. 
In contrast to the valence bands, the conduction bands localize more evenly on the \ce{Te} and \ce{Bi} sites, as can be seen from the partial $\sf DOS$ shown in Figure~\ref{fig:oc_spectral} (b).
Thus, for positive energies, increased tunneling into \ce{Bi} orbitals is outweighed by an decreased  tunneling into \ce{Te} orbitals leading to a small reduction of $\mathsf{Pk_2}$ and $\mathsf{Pk_3}$.
However, the ratio $\mathcal{R}$ is only $73\text{--}80\%$ for the peaks $\mathsf{Pk_2}$ and $\mathsf{Pk_3}$ at positive energies.

We base our analysis on the ratio $\mathcal{R}$ instead of the difference $\mathsf{dI/dV}(\mathrm{O}1,\mathrm{O}2)-\mathsf{dI/dV}(\ce{Te})$, which would be a physical quantity, but cannot be compared across different energies, due to exponential damping in the vacuum barrier. 
Correcting for the vacuum tunneling barrier is not straightforward \cite{Lang1986, Stroscio1986}.

The observations here are what we expect from the band structure plotted in Figure~\ref{fig:oc_spectral} (c). 
Bands which are localized mostly on the \ce{Te} orbitals appear red and the \ce{Bi} bands are plotted in blue. 
A dark tone indicates strong localization, a lighter shade indicates a weaker localization. The valence bands appear in a dark red, whereas the conduction bands are plotted in light blue color, confirming that the conduction bands localize more evenly on the \ce{Te} and \ce{Bi} than the valence bands.
The band inversion at $\sf \Gamma$ is clearly visible in the half-space band structure.

\subsection{Spatial Method}
\label{ssec:spatm}

\begin{figure}[h]
        \centering
        \includegraphics{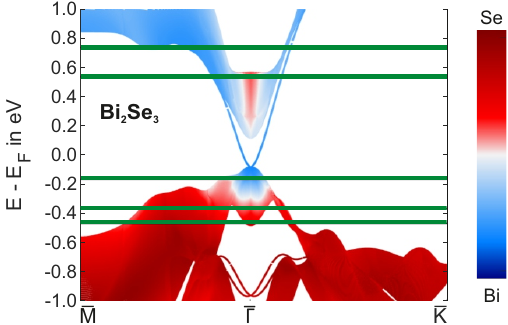}
        \caption{\label{fig:Bi2Se3_half-space_orbital} Orbital weighted half-space band structure of \ce{Bi2Se3}. The occupied bands are a linear combination of Wannier functions localized on the Selenium atomic positions, whereas the unoccupied bands are localized mostly on the Bismuth atoms. The band inversion at the $\overline{\sf \Gamma}$ point can be observed as the reverse ordering of the colors. Green lines indicate energies of the \dIdV maps.}
\end{figure}

In Ref.~\cite{Feenstra1987}, it was demonstrated that the electronic structure of \ce{GaAs} results in bias dependent imaging in constant current STM, where either the \ce{Ga} atoms or the \ce{As} atoms are selectively visible.
For negative bias voltages tunneling out of the occupied states make the \ce{As} appear bright, whereas for positive voltages the \ce{Ga} atoms are mapped as topographic protrusions.
Figure~\ref{fig:Bi2Se3_half-space_orbital} shows the half-space band structure of \ce{Bi2Se3} with color indicating the projection of the Bloch functions onto the Wannier functions localized on the \ce{Bi}, respectively \ce{Se}, atomic sites.
The half-space band structure indicates that the valence bands are primarily formed by orbitals of \ce{Se}, whereas the conduction bands consist mostly out of \ce{Bi} orbitals.
In this respect \ce{Bi2Se3} has a lot in common with other binary compound semiconductors.
However, the band inversion at $\sf \Gamma$, giving rise to the topological boundary mode, can be discerned clearly.

\begin{figure*}
        \centering
        \includegraphics{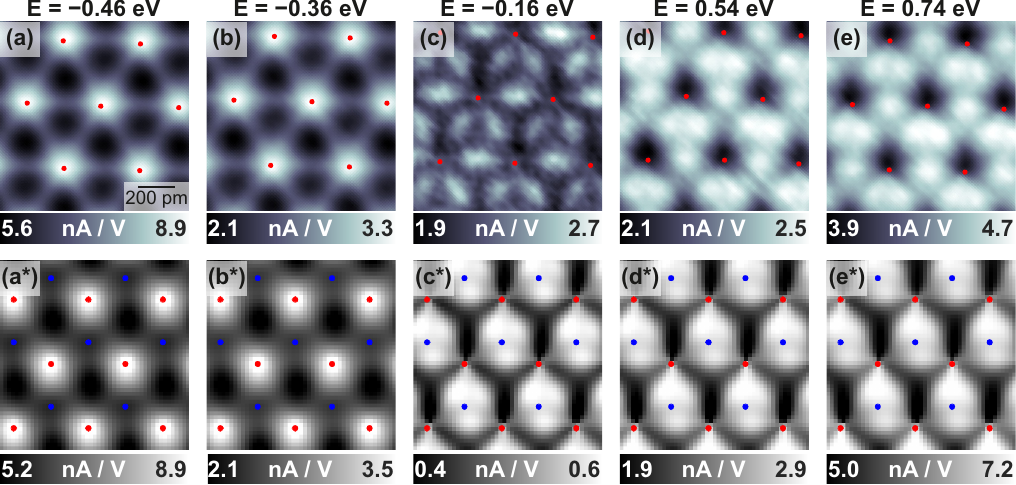}
        \caption{\label{fig:oc_spatial} (a) - (e) Experimentally obtained \dIdV maps for varying bias voltage between tip and sample.
        (a*) - (e*) Calculated \dIdV maps of the \ce{Bi2Se3} at various energies distributed throughout the valence and conduction bands. 
        Red dots indicate the lateral position of the \ce{Se} atoms and in the theoretical images blue dots indicate the lateral position of \ce{Bi} atoms.
        Lateral drift of the \ce{Se} atoms between the individual measurements is present in the experimental data.
       }
\end{figure*}

In this article we use a refinement of the method presented in Ref.~\cite{Feenstra1987}.
We measure the differential conductance $\mathsf{d}I/\mathsf{d}V$ instead of the total conductance.
This allows us to probe the electronic state at a certain energy $E$ instead of probing all states between $E$ and the Fermi energy $E_F$.
This is especially important, as in \ce{Bi2Se3} the lower conduction bands are occupied. 
As a result, the tunneling current is formed through a complex interplay of contributions from states strongly localized on \ce{Bi} atoms and those strongly localized on \ce{Se} atoms.
Green horizontal lines in Figure~\ref{fig:Bi2Se3_half-space_orbital} represent the energies at which \dIdV maps were measured.
The experimental \dIdV images at these energies are shown in Figure~\ref{fig:oc_spatial} (a) - (e).
The calculated \dIdV maps at the corresponding energies are shown in Figure~\ref{fig:oc_spatial} (a*) - (e*).
In the experimental images the lateral position of the \ce{Se} atoms is indicated by a red dot, which was determined by the maximum repulsive force in the simultaneously recorded frequency shift image \cite{SI, Liebig2022}.
Additionally, in the calculated images the lateral position of the \ce{Bi} atoms is plotted as a blue dot. 
Note that the data was measured using the same tip and sample as for the spectrum shown in Figure~\ref{fig:intro} (a).
We use the \dIdV spectrum obtained in the same height as the \dIdV maps to represent the data of Figure~\ref{fig:oc_spatial} in $\SI{}{nA/V}$, using the method described in the supplemental \cite{SI}. However, we need to account for $\SI{30}{pm}$ of vertical drift during the measurement of the \dIdV maps.

Taking the relaxation of the \ce{CO} molecule into account is necessary to achieve satisfactory agreement between experiment and theory in Figure~\ref{fig:oc_spatial}.
This relaxation was calculated using the PPM \cite{Hapala2014, Hapala2014_2}.
The bending of the \ce{CO} molecule leads to an transformation of the derivatives for describing the tunneling through $p$-type orbitals, as described in Subsection~\ref{ssec:cotc} and also a deflection of the position of the oxygen atom, which transmits most of the tunneling current.
A more detailed discussion can be found in the supplemental material \cite{SI}. 

For energies within the bulk valence band, \emph{i.e.}, Figure~\ref{fig:oc_spatial} (a)-(b) and (a*)-(b*), we observe the \ce{Se} atoms as bright features connected by fainter lines.
We note good agreement between experiment and theory.
The \ce{Se} atoms do not appear spherical, but rather as slightly irregular hexagons.
This deformation, as well as the lines can be traced back to the bending of the \ce{CO} molecule attached to the tip \cite{SI}.
In the calculated image the off sites at which the \ce{Bi} atoms are located are brighter than the empty off sites, which is not the case in the experimental data shown in Figure~\ref{fig:oc_spatial} (a)-(b).
The half-space band structure of \ce{Bi2Se3} indicates strong localization of the Bloch bands on the \ce{Se} atoms, which is in agreement with the experimental observation.
Our interpretation for negative energies agrees with previous experimental observations on \ce{Bi2Se3} and accompanying calculations of the $\sf LDOS$ from first-principles \cite{Urazhdin2004}.

In Figure~\ref{fig:oc_spatial} (c) and (c*), close to the VBM, we observe lines or elongated maxima between the \ce{Se} atoms in both the experimental and calculated images.
In contrast to Panels (a), (a*), (b) and (b*), we no longer observe the \ce{Se} atoms as bright features.
Furthermore, in the theoretical data of Figure~\ref{fig:oc_spatial} (c*), the empty off sites appear as dark $\sf Y$ shaped feature.
Centered on the \ce{Bi} sites there is a \rotatebox[origin=c]{180}{\textsf{Y}} shaped region, with three \ce{Se} atoms at the ends. This \rotatebox[origin=c]{180}{\textsf{Y}} shape is dimmer than the elongated maxima along the direct connecting lines between each pair of \ce{Se} atoms.
In the experimental data shown in Figure~\ref{fig:oc_spatial} (c) we see similar tendencies. However, the \rotatebox[origin=c]{180}{\textsf{Y}} shaped feature is only slightly brighter than the \textsf{Y} centered on the other off site. 
The elongated maxima between the pairs of \ce{Se} atoms are much brighter than both the \textsf{Y} and \rotatebox[origin=c]{180}{\textsf{Y}} shape.
An energy dependent variation of the mix of tip orbitals might explain this variation in the brightness.
The half-space band structure in Figure~\ref{fig:Bi2Se3_half-space_orbital} indicates that an energy of $E = \SI{-0.16}{eV}$ corresponds to the valence band edge, where the bands are inverted, explaining the absence of electronic states localized on the \ce{Se} atoms that contribute to the \dIdV signal.
As the \ce{Bi} atoms are located in the second atomic layer, the Bloch states localized on the \ce{Bi} are not directly observed as bright features.


For positive energies, \emph{i.e.}, Panels (d), (d*), (e) and (e*) we observe bright features on the upwards facing off sites, corresponding to the \ce{Bi} sites.
These bright features consist again of elongated features with a high strength of the \dIdV signal connecting pairs of \ce{Se} atoms and dimmer \rotatebox[origin=c]{180}{\textsf{Y}} shaped regions centered on the upwards facing off sites.
On the empty offsite we observe dark \textsf{Y} shaped features connecting the \ce{Se} atoms.
In the experimental images the \ce{Se} atoms appear as dark images, whereas in the theoretical images we see sharp, bright lines at the \ce{Se} atoms.
This tendency might be explained by the fact that we do not consider the convolution of tip and sample wavefunction, as shown in Equation~\eqref{M}, but applied Chens derivative rule.
The calculated images might not be as smooth as the experimental images, due to not considering this convolution.
From the half-space band structure in Figure~\ref{fig:Bi2Se3_half-space_orbital} we might have expected the \ce{Se} atoms to appear bright in Figure~\ref{fig:oc_spatial} (d) due to the band inversion. However, the band inversion at the conduction band edge appears only in in light red. Thus the color map of Figure~\ref{fig:Bi2Se3_half-space_orbital} indicates only a weak localization on the \ce{Se} atoms. Furthermore, the conclusion in Subsection~\ref{Bi2Se3_STS} demonstrated, that the states at conduction band edge close to $\overline{\sf{\Gamma}}$ have a negligible contribution to the \dIdV signal, opposed to the states farther away from $\overline{\sf{\Gamma}}$ in this energy region.
The states at the conduction band edge farther away from $\overline{\sf{\Gamma}}$, that contribute strongly to the \dIdV signal, are however again localized predominantly on the \ce{Bi} atoms, explaining why we do not observe an increased \dIdV signal above the \ce{Se} atoms due the band inversion at the conduction band edge.

In the model presented here, there are possible improvements.
We note, that in all calculated images of Figure~\ref{fig:oc_spatial} we assumed that the linear combination of wavefunctions at the tip does not depend on the energy $E$.
Making this linear combination energy dependent might increase the agreement between experiment and theory, but introduces additional parameters.
Another possible improvement is including bias and current dependent tip-sample forces in the PPM calculation.
This includes electrostatic forces or current-induced electrostatic forces~\cite{Weymouth2011}, which might add to the deflection of the \ce{CO} molecule.
\vspace{-2mm}
\section{Outlook}
\label{sec:out}
\vspace{-1mm}
Incorporating defects into the tight-binding model, as outlined in Ref.~\cite{Liu2013}, allows for the study of point defect signatures using the presented method. Comparing theoretical and experimental \dIdV spectra enables precise calibration of the energy scale, while the real-space sensitivity would allow to selectively probe for defect states on the atomic scale \cite{Queiroz2024}.
Furthermore it is conceivable to adapt our calculations further and study the evolution of the band structure in Moir\'e patterned samples \cite{Schouteden2016} or the changes induced by external magnetic fields \cite{Peng2010, Yeping2012, Hanaguri2010}.
Future applications of the methods presented here include observing local phase transitions, possibly induced by strain \cite{Liu2014} or  by substitutional atoms \cite{SanchezBarriga2018}. 
From a fundamental and theoretical perspective the robustness of the boundary mode against crystal defects and changes in geometry of the crystal boundary has come into focus \cite{Ludewig2022_cob, Ewert2019}. 
The derivation of the topological invariants by means of Bloch theory relies heavily on translational invariance. 
Consequently, it is not clear how the topological classification of materials extends to disordered materials.
We believe that the experimental results presented here establish that combined STM and AFM is an important tool for research in this area, as it is able to simultaneously characterize samples structurally and electronically from the atomic scale upward.


\begin{acknowledgments}
We are thankful for very interesting discussions with Christoph Friedrich and his support with the use of the code SPEX. 
We are grateful to Stefan Blügel for establishing the contact between experiment and theory.
We thank Marco Weiss for implementing a LabView program for automated measurement of \dIdV maps.
For the first-principles calculations, the computational resources were provided by the Dutch national e-infrastructure with the support of the SURF Cooperative under Grants No. EINF-5380 and EINF-10348. We acknowledge the ﬁnancial support of the Deutsche
Forschungsgemeinschaft through Project ID No. 422
314695032-SFB1277 (Subprojects No. A01, No. A08 and No. B02). 
We are thankful to Jay Weymouth and Fabian Stilp for careful proofreading.
\end{acknowledgments}



%

\end{document}